\documentclass[fleqn,usenatbib]{mnras}

\usepackage{newtxtext,newtxmath}
\usepackage{rotating}

\usepackage[T1]{fontenc}
\usepackage{ae,aecompl}
\usepackage{amsmath} 
\newcommand{\angstrom}{\textup{\AA}}

\usepackage{graphicx}	
\usepackage{amsmath}	
\usepackage{gensymb}
\usepackage[flushleft]{threeparttable}
\usepackage[usenames, dvipsnames]{color}

\def\gax{\mathrel{\raise.3ex\hbox{$>$}\mkern-14mu\lower0.6ex\hbox{$\sim$}}}
\def\lax{\mathrel{\raise.3ex\hbox{$<$}\mkern-14mu\lower0.6ex\hbox{$\sim$}}}
\def\gtorder{\mathrel{\raise.3ex\hbox{$>$}\mkern-14mu
             \lower0.6ex\hbox{$\sim$}}}
\def\ltorder{\mathrel{\raise.3ex\hbox{$<$}\mkern-14mu
             \lower0.6ex\hbox{$\sim$}}}
             
\title[MACHO 80.7443.1718: the most extreme heartbeat star]{The Loudest Stellar Heartbeat: Characterizing the most extreme amplitude heartbeat star system}

\author[T. Jayasinghe et al.]{T. Jayasinghe$^{1,2}$\thanks{E-mail: jayasinghearachchilage.1@osu.edu},
C. S. Kochanek$^{1,2}$,
J. Strader$^{3}$,
K. Z. Stanek$^{1,2}$,
P. J. Vallely$^{1,2}$,
\newauthor 
Todd A. Thompson$^{1,2}$,
J. T. Hinkle$^{4}$,
B. J. Shappee$^{4}$,
A. K. Dupree$^{5}$,
\newauthor 
K. Auchettl$^{6,7,8}$,
L. Chomiuk$^{3}$,
E. Aydi$^{3}$,
K. Dage$^{9,10,3}$,
A. Hughes$^{11}$,
\newauthor
L. Shishkovsky$^{3}$,
K. V. Sokolovsky$^{3,12}$,
S. Swihart$^{3,13}$,
K. T. Voggel$^{14}$,
I. B. Thompson$^{15}$
\\
$^{1}$Department of Astronomy, The Ohio State University, 140 West 18th Avenue, Columbus, OH 43210, USA\\
$^{2}$Center for Cosmology and Astroparticle Physics, The Ohio State University, 191 W. Woodruff Avenue, Columbus, OH 43210, USA\\
$^{3}$Center for Data Intensive and Time Domain Astronomy, Department of Physics and Astronomy, Michigan State University, East Lansing MI 48824, USA\\
$^{4}$Institute for Astronomy, University of Hawaii, 2680 Woodlawn Drive, Honolulu, HI 96822, USA\\
$^{5}$Center for Astrophysics, Harvard \& Smithsonian, 60 Garden Street, MS-15, Cambridge, MA 02138, USA\\
$^{6}$School of Physics, The University of Melbourne, Parkville, VIC 3010, Australia\\
$^{7}$ARC Centre of Excellence for All Sky Astrophysics in 3 Dimensions (ASTRO 3D)\\
$^{8}$Department of Astronomy and Astrophysics, University of California, Santa Cruz, CA 95064, USA\\
$^{9}$Department of Physics, McGill University, 3600 University Street, Montr\'eal, QC H3A 2T8, Canada\\
$^{10}$McGill Space Institute, McGill University, 3550 University Street, Montr\'eal, QC H3A 2A7, Canada\\
$^{11}$Department of Astronomy/Steward Observatory, 933 North Cherry Avenue, Rm. N204, Tucson, AZ 85721-0065, USA\\
$^{12}$Sternberg Astronomical Institute, Moscow State University, Universitetskii~pr.~13, 119992~Moscow, Russia\\
$^{13}$National Research Council Research Associate, National Academy of Sciences, Washington, DC 20001, USA,\\
resident at Naval Research Laboratory, Washington, DC 20375, USA\\
$^{14}$Universite de Strasbourg, CNRS, Observatoire astronomique de Strasbourg, UMR 7550, 67000 Strasbourg, France\\
$^{15}$Carnegie Observatories, 813 Santa Barbara Street, Pasadena, CA 91101-1292, USA\\
}

\date{Accepted XXX. Received YYY; in original form ZZZ}

\pubyear{2020}
 
\begin{document}
\label{firstpage}
\pagerange{\pageref{firstpage}--\pageref{lastpage}}
\maketitle

\begin{abstract}
We characterize the extreme heartbeat star system MACHO 80.7443.1718 in the LMC using \textit{TESS} photometry and spectroscopic observations from the Magellan Inamori Kyocera Echelle (MIKE) and SOAR Goodman spectographs. MACHO 80.7443.1718 was first identified as a heartbeat star system in the All-Sky Automated Survey for SuperNovae (ASAS-SN) with $P_{\rm orb}=32.836\pm0.008\,{\rm d}$. MACHO 80.7443.1718 is a young (${\sim}6$~Myr), massive binary, composed of a B0 Iae supergiant with $M_1 \simeq 35 M_\odot$ and an O9.5V secondary with $M_2 \simeq 16 M_\odot$ on an eccentric ($e=0.51\pm0.03$) orbit. In addition to having the largest variability amplitude amongst all known heartbeats stars, MACHO 80.7443.1718 is also one of the most massive heartbeat stars yet discovered. The B[e] supergiant has Balmer emission lines and permitted/forbidden metallic emission lines associated with a circumstellar disk. The disk rapidly dissipates at periastron which could indicate mass transfer to the secondary, but re-emerges immediately following periastron passage. MACHO 80.7443.1718 also shows tidally excited oscillations at the $N=25$ and $N=41$ orbital harmonics and has a rotational period of 4.4 d. 
\end{abstract}




\section{Introduction}

Heartbeat stars are binaries with short period ($P\lesssim1$\,yr), eccentric ($e\gtrsim0.3$) orbits. The light curves of heartbeat stars are defined by oscillations outside of periastron combined with a brief, high amplitude ellipsoidal variations at periastron that gives rise to a unique ``heartbeat'' signature. Heartbeat stars were first discovered and characterized using data from the \textit{Kepler} space telescope \citep{Borucki,2012ApJ...753...86T,2016AJ....151...68K} and its follow-up mission K2 \citep{2014PASP..126..398H}. \textit{Kepler} identified over 170 heartbeat stars \citep{2016AJ....151...68K}. The photometric modulations of these systems are dominated by the effects of tidal distortion, reflection and Doppler beaming close to periastron \citep{1995ApJ...449..294K,2017MNRAS.472.1538F}. The variability amplitude of most heartbeat stars is very small ($\lesssim 1$ mmag; \citealt{2016AJ....151...68K,2018MNRAS.473.5165H}). 

Heartbeat stars show tidally excited oscillations (TEOs) at exact integer multiples of the orbital frequency \citep{2017MNRAS.472.1538F} driven by the tidal forcing at periastron. The largest amplitude TEOs are resonances between harmonics of the orbital frequency and the normal mode frequencies of the star \citep{2017MNRAS.472.1538F,2020arXiv200901851C}. Most known heartbeat stars are relatively low-mass A and F type stars. However, several massive OB-type heartbeat stars have recently been discovered. For example, $\iota$ Ori is a massive heartbeat star system consisting of a O9 III primary and a B1 III-IV companion discovered by BRITE \citep{2017MNRAS.467.2494P} and $\epsilon$ Lupi consists of two early-type main sequence B stars \citep{2019MNRAS.488...64P}. Recently, \citet{Kolaczek-Szymanski2021} identified 20 massive heartbeat stars using data from the Transiting Exoplanet Survey Satellite (\textit{TESS}; \citealt{2015JATIS...1a4003R}). Heartbeat stars with TEOs are useful laboratories to study equilibrium and dynamical tides \citep{2020ApJ...888...95G}, and massive heartbeat systems can be used to study these processes in the context of binary systems that are the progenitors for compact objects. 

MACHO 80.7443.1718 with J2000 coordinates $(\alpha,\delta)=(81.601924^\circ,-68.784706^\circ)$, was initially classified as an eclipsing binary in the Large Magellanic Cloud (LMC) with a period of $P{\sim}32.8\;$d by the MACHO survey \citep{1997ApJ...486..697A}. In our previous work \citep{2019MNRAS.489.4705J}, we identified it as a heartbeat star (ASASSN-V J052624.38-684705.6) using data from The All-Sky Automated Survey for SuperNovae (ASAS-SN, \citealt{2014ApJ...788...48S, 2017PASP..129j4502K,2018MNRAS.477.3145J,2019MNRAS.486.1907J}) and \textit{TESS}. MACHO 80.7443.1718 is a massive star that is part of the LH58 OB association in the LMC, northwest of 30 Doradus. An archival spectrum classified it as a B0.5 Ib/II \citep{1994AJ....108.1256G}, evolved blue star with $U-B=-0.84$ mag, and $B-V=0.11$\,mag \citep{2002ApJS..141...81M}. We found that MACHO 80.7443.1718 displayed the largest known heartbeat flux variations, with a peak-to-peak variability of ${\sim}40\%$ at periastron and oscillation amplitudes of ${\sim}10\%$ due to TEOs outside periastron.  

Here we characterize this extreme amplitude heartbeat star using \textit{TESS} photometry and spectroscopic observations from the Magellan/MIKE and SOAR/Goodman spectrographs. We discuss the TESS observations and spectroscopic followup in Section 2. In Section 3, we characterize the binary orbit, stellar parameters, and TEOs of this binary system. In Section 4, we discuss the implications of our observations of this unique heartbeat star system. We present a summary of our results in Section 5.

\section{Observations}

\subsection{Photometry}

MACHO 80.7443.1718 lies in the Southern \textit{TESS} CVZ which allowed us to extract TESS light curves for both Sectors 1 and 2 in \citet{2019MNRAS.489.4705J}. Here, we extend the TESS photometry using data from sectors 1-7, 9-10, 12-13, 27 and 29-34. We analyzed the \textit{TESS} data using an image subtraction pipeline derived from that used to process ASAS-SN data which has been optimized for use with the \textit{TESS} FFIs \citep{Vallely2020}.
This pipeline is based on the ISIS package \citep{1998ApJ...503..325A,2000A&AS..144..363A}, and a detailed description of the \textit{TESS}-specific corrective procedures can be found in \citet{Vallely2020}.

\begin{figure*}

	\includegraphics[width=1.05\textwidth]{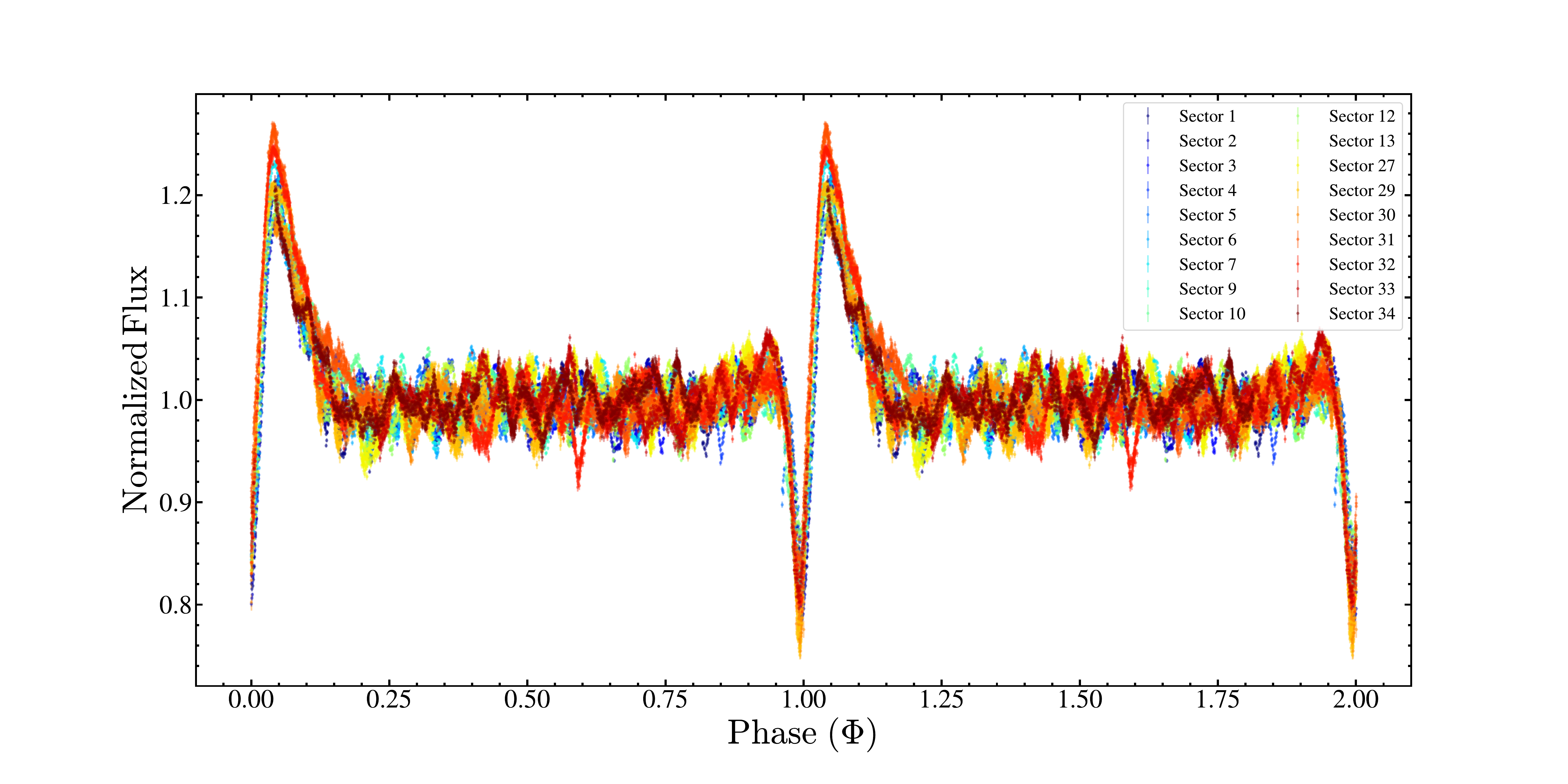}
    \caption{The phased \textit{TESS} light curve for MACHO 80.7443.1718. The points are colored by the sector.}
    \label{fig:fig1}

\end{figure*}

While this pipeline produces excellent differential flux light curves, the large pixel scale of \textit{TESS} makes it difficult to obtain reliable measurements of the reference flux of a given source. Here we have estimated the reference flux using the \verb"ticgen" software package \citep{ticgen,2018AJ....156..102S}. The \textit{TESS}-band magnitude estimates from \verb"ticgen" were converted into fluxes using an instrumental zero point of 20.44 electrons per second in the FFIs, based on the values provided in the TESS Instrument Handbook. Flux was then added to the raw differential light curves such that the median of each sector's observations matched the estimated reference value. This allowed us to produce the normalized flux light curves shown in Figure~\ref{fig:fig1}. The light curves do not include epochs where the observations were compromised by scattered light artifacts from the Earth or Moon.

From the combined \textit{TESS} light curve, we derive an orbital period of\begin{equation}
    \rm P_{\rm orb, TESS}=32.8194\pm0.0105\,{\rm d}, 
	\label{eq:period}
\end{equation} which is consistent with the period of \begin{equation}
    \rm P_{\rm orb}=32.83627\pm0.00846\,{\rm d}, 
	\label{eq:asassnperiod}
\end{equation} derived from the ASAS-SN light curve. For the remainder of this work, we adopt the ASAS-SN orbital period.
We use the periastron ephemeris derived from the \textit{TESS} data of,
\begin{equation}
    \rm {\rm BJD}_{\rm periastron}= 2458373.61518 +32.83627 \times E\,,
	\label{eq:ephemtess}
\end{equation}
where the epoch $\rm E$ is the number of orbits since the time of minimum. Figure \ref{fig:fig1} shows the phased \textit{TESS} light curve for MACHO 80.7443.1718. The strong ellipsoidal variation and TEOs are clearly resolved in the \textit{TESS} data. Following the photometric maximum, we see significant variations in the descending branch of the ellipsoidal variation signal ($0.05<\Phi<0.15$). The inclusion of the 16 additional \textit{TESS} sectors significantly improves our previous analysis of the TEOs (see $\S3.4$).

We also retrieve 23 archival photometric measurements spanning $22.2\mu$m ($W4$) through the $U$-band (Table \ref{tab:phot}). In addition, we obtained archival Swift UVOT \citep{roming05} images in the $UVW1$ (2600 \AA), $UVM2$ (2246 \AA), and $UVW2$ (1928 \AA) bands \citep{poole08} from the LMC Survey \citep{Hagen2017} pointings 148 and 16. Each epoch of UVOT data includes 2 observations per filter, which we combined using the \texttt{uvotimsum} package. We then used \texttt{uvotsource} to extract source counts using a 6\farcs{0} radius aperture centered on the star. We measured the background counts using a source-free region with radius of $\sim$18\farcs{0}. Using the most recent calibrations \citep{poole08, breeveld10}, we converted the UVOT counts into fluxes and magnitudes. We combined all 4 epochs with a weighted average and summed the uncertainties on the individual epochs with the standard deviation in quadrature to obtain our error estimate. This photometry is used for the SED models in $\S3.2$.

\begin{table}
	\centering
	\caption{Multi-band photometry measurements for MACHO 80.7443.1718}
	\label{tab:phot}
\begin{tabular}{rrrr}
		\hline
		 Magnitude & $\sigma$ & Filter & Reference\\
		\hline

14.24 & 0.04 &  UVM2 & This work \\
14.22 & 0.03 &  UVW2 &  This work \\
13.92 & 0.04 &  UVW1 & This work \\
12.83 & 0.10 & U   &\citet{2002ApJS..141...81M} \\
12.630 & 0.038 & U & \citet{2004AJ....128.1606Z} \\
13.67 & 0.10 & B  & \citet{2002ApJS..141...81M} \\
13.617 & 0.106 & B & \citet{2004AJ....128.1606Z} \\
13.608 & 0.262 & V & \citet{2004AJ....128.1606Z} \\
13.56 & 0.10 & V  & \citet{2002ApJS..141...81M} \\
13.43 & 0.10 & R  & \citet{2002ApJS..141...81M} \\
13.283 & 0.079 & I & \citet{2004AJ....128.1606Z} \\
13.255 & 0.007 &  I &  \citet{cioni} \\
13.020 & 0.022 &  J & \citet{2003yCat.2246....0C} \\
12.978 & 0.023 & J & \citet{cioni} \\ 
12.833 & 0.022 &  H & \citet{2003yCat.2246....0C} \\
12.734 & 0.030 & $K_s$ & \citet{2003yCat.2246....0C} \\
12.411 & 0.033 &  [3.6] & \citet{2006AJ....132.2268M} \\
12.300 & 0.030 &  [4.5] & \citet{2006AJ....132.2268M} \\
12.135 & 0.067 &  [5.8] & \citet{2006AJ....132.2268M} \\
12.416 & 0.024 &  W1 & \citet{2010AJ....140.1868W} \\
12.337 & 0.022 &  W2 & \citet{2010AJ....140.1868W} \\
9.901 & 0.071 &  W3 & \citet{2010AJ....140.1868W} \\
6.197 & 0.092 &  W4 & \citet{2010AJ....140.1868W} \\
\hline
\end{tabular}
\end{table}

\subsection{Optical Spectroscopy}
\begin{figure*}

	\includegraphics[width=\textwidth]{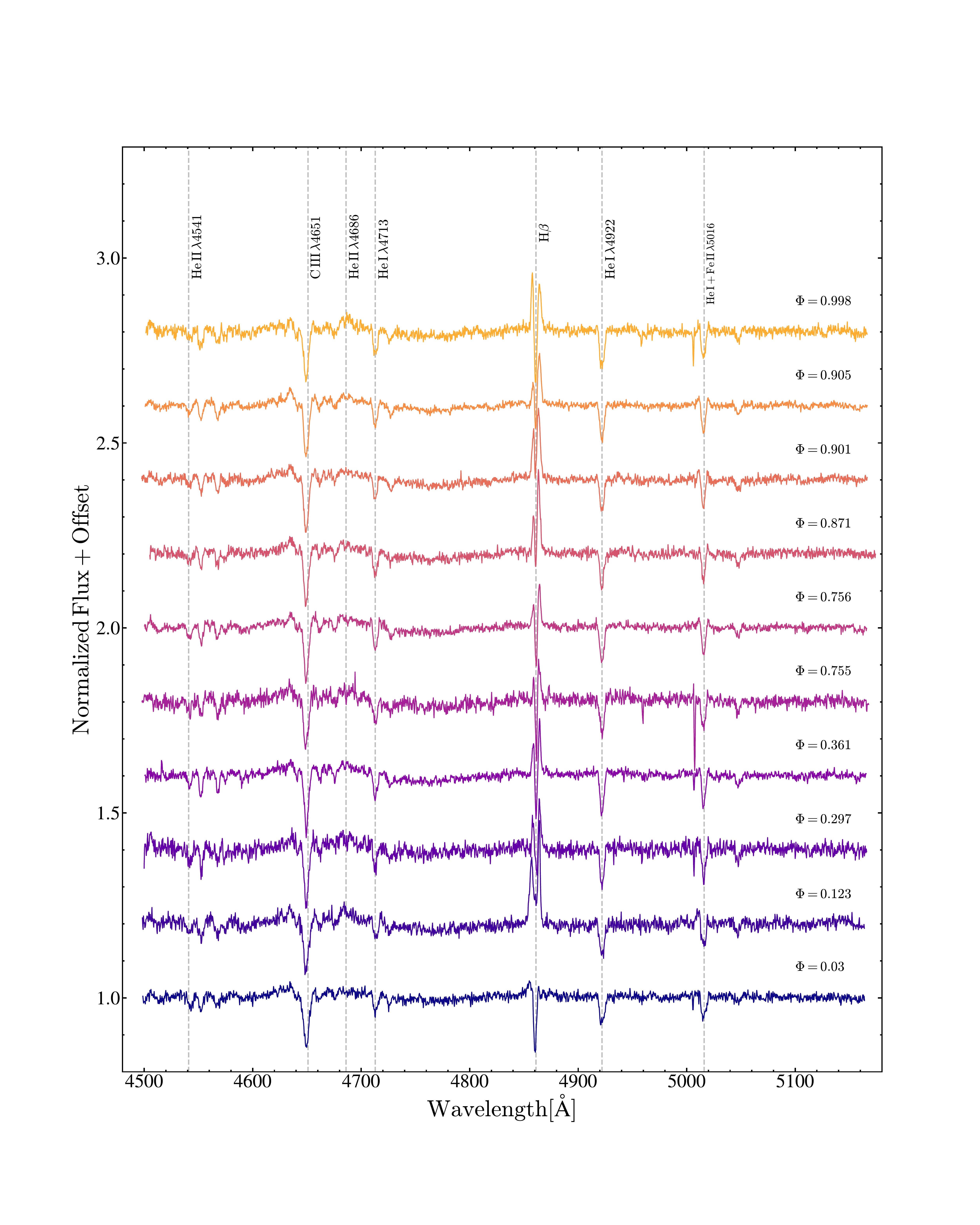}
    \caption{SOAR spectra for MACHO 80.7443.1718 sorted by the orbital phase $\Phi$, where $\Phi=0$ corresponds to periastron.}
    \label{fig:fig2}

\end{figure*}

We obtained 17 epochs of spectroscopy: 4 with Magellan/MIKE \citep{2003SPIE.4841.1694B}, and 13 with SOAR/Goodman \citep{2004SPIE.5492..331C}, between 2019 Jan 6 and 2019 Aug 6. The first and third MIKE epochs were obtained with a 0\farcs{7} slit, while the other two epochs used a 1\farcs{0} slit. The resolution of the spectra in the region of interest (see below) was about 0.12 \AA\ and 0.16 \AA, respectively, for the two slit widths. Data were obtained for both the blue and red sides of the instrument; here, only data from the blue camera are used. The MIKE data were reduced using \emph{CarPy} \citep{2000ApJ...531..184K,2003PASP..115..688K}. The SOAR spectra were taken with a 2100 lines mm$^{-1}$ grating and a 0\farcs{95} slit, giving a resolution of 1.0 \AA\ over the wavelength range $\sim 4500$--5170 \AA. The exposure time was 1200 s per epoch. The data were reduced and optimally extracted in the standard manner. For both datasets, we report the mid-exposure observation times as Barycentric Julian Dates (BJD) on the TDB system \citep{2010PASP..122..935E}. Figure \ref{fig:fig2} illustrates the SOAR spectra (excluding two epochs with low SNR). We see many emission lines, including the Balmer $\rm H\beta$ line, suggesting that this source is likely a Be star with a circumstellar disk.

In the majority of the spectra, most of the strong absorption lines are contaminated by emission and hence unusable for radial velocities. The strongest line without evidence for emission is the \ion{He}{i} line at $4922 \angstrom$. For the MIKE spectra, we used this absorption line to determine velocities, using a spectrum of the B1 III star HD 68761 from the UVES POP library \citep{2003Msngr.114...10B} as a template. For the SOAR data, we used the highest signal-to-noise SOAR spectrum (that of 2019 Mar 24) as a template and obtained radial velocities of the other SOAR spectra through cross-correlation with this template, again in the region of the $4922\angstrom$ line. The velocity of the template was determined through cross-correlation with a spectrum of the bright B0 III star HD 114122 taken with the same instrumental setup. The barycentric radial velocities of all the spectra are given in Table \ref{tab:rv}. Note that, due to the rapid rotation of the star and the resulting broad lines, the MIKE spectra do not have substantially smaller velocity uncertainties than the SOAR spectra. 

\begin{table*}
	\centering
	\caption{Radial velocity (RV) measurements for MACHO 80.7443.1718 from SOAR and MIKE.}
	\label{tab:rv}
\begin{tabular}{rrrrrr}
		\hline
		 BJD & Date & Phase ($\Phi)$ & RV ($\rm km s^{-1}$) & $\sigma_{RV}$ ($\rm km s^{-1}$) & Instrument\\
		\hline
2458489.6195626 & 2019-01-06 & 0.533 & 271.0 & 5.8 & MIKE \\
2458498.5758614 & 2019-01-15 & 0.806 & 251.3 & 5.2 & MIKE \\
2458517.5984611 & 2019-02-03 & 0.385 & 294.7 & 6.2 & MIKE \\
2458491.6776199 & 2019-01-08 & 0.595 & 254.2 & 9.8 & SOAR \\
2458525.5646330 & 2019-02-11 & 0.627 & 256.2 & 12.8 & SOAR \\
2458547.5450865 & 2019-03-05 & 0.297 & 301.2 & 4.9 & SOAR \\
2458567.4966222 & 2019-03-24 & 0.905 & 245.3 & 4.0 & SOAR \\
2458579.5027003 & 2019-04-06 & 0.270 & 315.2 & 6.5 & MIKE \\
2458581.4869998 & 2019-04-07 & 0.331 & 306.8 & 4.5 & SOAR \\
2458582.4874062 & 2019-04-08 & 0.361 & 294.7 & 4.8 & SOAR \\
2458595.4763313 & 2019-04-21 & 0.756 & 247.8 & 4.4 & SOAR \\
2458604.4649556 & 2019-04-30 & 0.030 & 350.9 & 6.9 & SOAR \\
2458607.5096233 & 2019-05-04 & 0.123 & 360.9 & 7.3 & SOAR \\
2458664.9182869 & 2019-06-30 & 0.871 & 237.0 & 6.0 & SOAR \\
2458665.8982865 & 2019-07-01 & 0.901 & 252.8 & 5.3 & SOAR \\
2458693.9235514 & 2019-07-29 & 0.755 & 251.3 & 6.4 & SOAR \\
2458701.9156224 & 2019-08-06 & 0.998 & 313.8 & 5.5 & SOAR \\

\hline
\end{tabular}
\end{table*}

\section{Results}

\subsection{Spectral classification}

 Using a low-resolution spectrum, MACHO 80.7443.1718 was previously classified as a B0.5 Ib/II star by \citet{1994AJ....108.1256G}. As we previously noted in $\S 2$, MACHO 80.7443.1718 is an emission line star with significant continuum emission from a disk that complicates analysis of the spectra. To estimate the spectral type, we calculate the equivalent widths and their ratios in the MIKE spectra for various lines that are typically used to assign spectral types for OB stars \citep{1971ApJ...170..325C,1988A&AS...76..427M, 2011ApJS..193...24S,2014ApJS..211...10S,2018A&A...616A.135M}. These results are summarized in Table \ref{tab:eqw}.
 
 Based on the classical spectral type criterion for OB stars using the ratio $\rm \ion {He}{i} \, 4471$/$\rm \ion {He}{ii} \, 4542$, MACHO 80.7443.1718 has a spectral type later than O9.5 \citep{1971ApJ...170..325C,1988A&AS...76..427M}. The ratio $\rm \ion {He}{i} \, 4144$/$\rm \ion {He}{ii} \, 4200=1.66\pm0.50$ is very similar to O9.7 stars with $\rm \ion {He}{i} \, 4144$/$\rm \ion {He}{ii} \, 4200=1.98\pm1.00$ \citep{2018A&A...616A.135M}. The average ratio of  $\rm \ion {He}{i} \, 4388$/$\rm \ion {He}{ii} \, 4542=3.37\pm0.52$ is somewhat different from O9.7 stars ($2.45\pm0.99$) but falls within the reported dispersion in \citet{2018A&A...616A.135M}. The definition of the O9.7 spectral type is based on the ratio $\rm \ion {Si}{iii} \, \lambda 4552$/$\rm \ion {He}{ii} \, \lambda 4542{\sim}1.0$ \citep{2011ApJS..193...24S}. However, even though MACHO 80.7443.1718 has $\rm \ion {Si}{iii} \, \lambda 4552$/$\rm \ion {He}{ii} \, \lambda 4542{\sim}1.6$, which is larger than this canonical definition, there are a few stars with O9.7 spectral types that have values close to ${\sim}2$ \citep{2018A&A...616A.135M}. It is also possible that MACHO 80.7443.1718 is an early B0/B0.2 star. Based on this analysis, MACHO 80.7443.1718 likely has a spectral type in the range O9.7--B0.2. Here, we will adopt the B0 spectral type for simplicity.
 
 The luminosity classes of late-O stars can be determined using the ratios $\rm \ion {Si}{iv} \,  4089$/$\rm \ion {He}{i}\, 4026$ and $\rm \ion {He}{ii} \, 4686$/$\rm \ion {He}{i} \,  4713$ which differ significantly for dwarfs and supergiants \citep{2011ApJS..193...24S,2014ApJS..211...10S,2018A&A...616A.135M}. Given the uncertainty in our assigned spectral type, we will utilize these criteria to assign a luminosity class even though MACHO 80.7443.1718 was assigned an early B0 spectral type. MACHO 80.7443.1718 has $\rm \ion {Si}{iv} \,  4089$/$\rm \ion {He}{i}\,  4026=1.20\pm0.19$ which is most comparable to Ia supergiants that have $\rm \ion {Si}{iv} \,4089$/$\rm \ion {He}{i}\,  4026=1.07\pm0.20$ \citep{2018A&A...616A.135M}. I/Iab stars have  $\rm \ion {Si}{iv} \,4089$/$\rm \ion {He}{i}\,  4026=0.85\pm0.11$ \citep{2018A&A...616A.135M}. Similarly, MACHO 80.7443.1718 has $\rm \ion {He}{ii} \,4686$/$\rm \ion {He}{i} \, 4713=0.12\pm0.07$ which is again comparable to Ia supergiants with $\rm \ion {He}{ii} \,4686$/$\rm \ion {He}{i} \,  4713=0.31\pm0.10$ \citep{2018A&A...616A.135M}. In comparison, I/Iab stars have  $\rm \ion {He}{ii} \,4686$/$\rm \ion {He}{i} \,  4713=0.90\pm0.30$ \citep{2018A&A...616A.135M}.
 
Based on this analysis, we assign the B0 Iae spectral type to the primary component of this heartbeat star system. Even though the assignment of a spectral type to this system is complicated by spectral variability, continuum emission from the disk, and the contribution of the companion star, our spectral classification does not differ significantly from that of \citet{1994AJ....108.1256G}. 

\begin{figure*}

	\includegraphics[width=\textwidth]{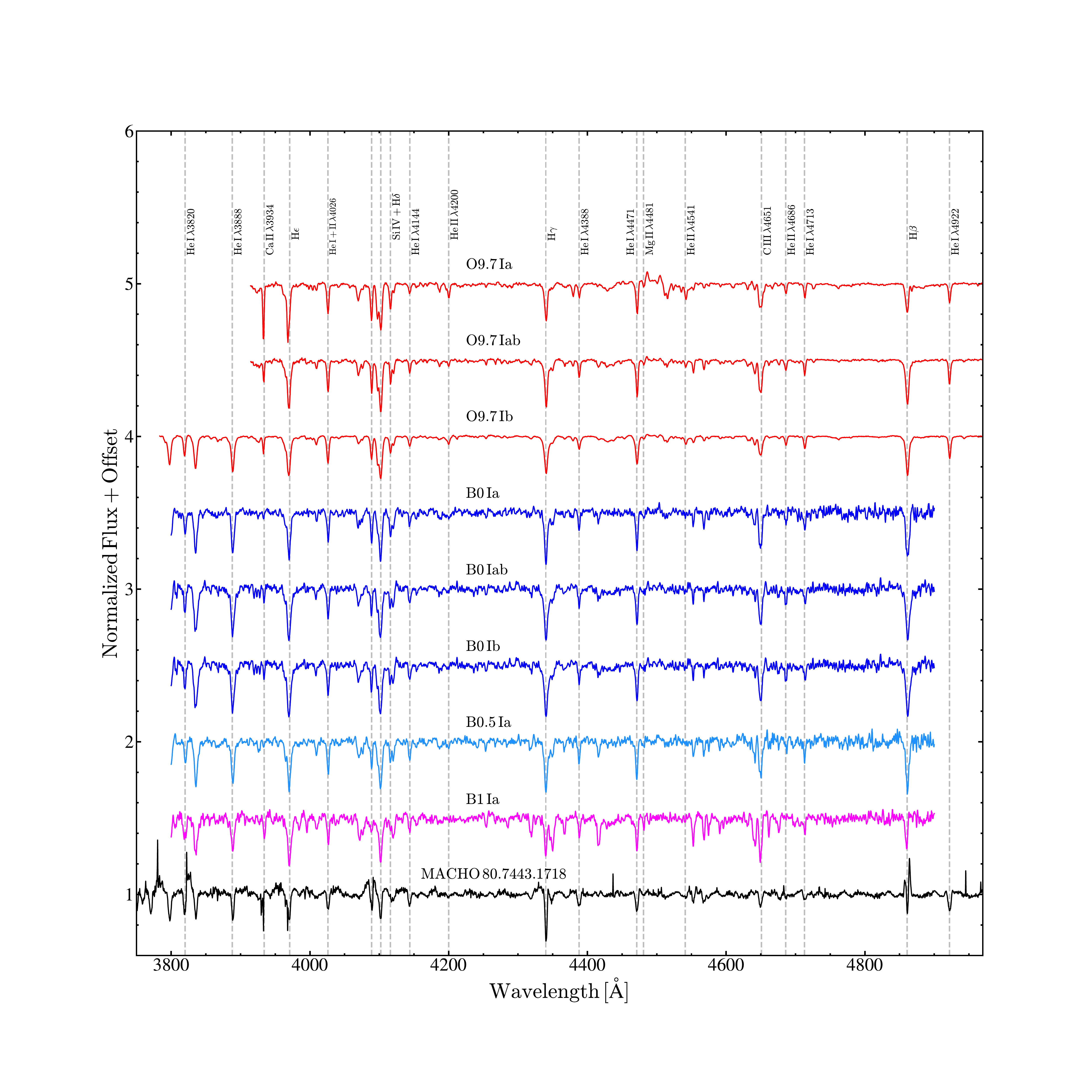}
	\vspace{-1.5cm}	
    \caption{Comparison of various spectral type standards from \citet{2013msao.confE.198M} and \citet{1990PASP..102..379W} with the MIKE spectrum at phase $\Phi=0.270$ (black).}
    \label{fig:fig3}

\end{figure*}

In Figure \ref{fig:fig3}, we compare the MIKE spectrum taken on UT 2019-04-06 ($\Phi=0.270$) with spectra of standard late-O and early-B supergiants. The O-star spectra are obtained from the Galactic O-star catalog \citep{2013msao.confE.198M} and the B-star spectra are obtained from \citet{1990PASP..102..379W}. While there are differences between the spectra, the assigned B0 spectral type is a reasonable match to the templates. Visually, based on the temperature sensitive $\rm \ion {Mg}{ii} \,4481$ doublet, MACHO 80.7443.1718 should have a spectral type earlier than B1. We measure an average equivalent width of $\rm EW(\ion {Mg}{ii} \,4481)=57\pm7\, m\angstrom$ which is consistent with a late-O or early-B spectral type \citep{1997A&A...317..871L}. 
 
 \begin{table}
	\centering
	\caption{The average equivalent width ratios and their dispersion calculated using the MIKE spectra.}
	\label{tab:eqw}
\begin{tabular}{cr}
		\hline
		 EW Ratio & Measurement\\
		\hline
		$\rm \ion {He}{i}\, \lambda 4471$/$\rm \ion {He}{ii} \, \lambda 4542$ & $5.16\pm0.71$ \\		
		$\rm \ion {He}{i} \, \lambda 4144$/$\rm \ion {He}{ii} \, \lambda 4200$ & $1.66\pm0.50$ \\
		$\rm \ion {He}{i} \,\lambda 4388$/$\rm \ion {He}{ii} \, \lambda 4542$ & $3.37\pm0.52$ \\
		$\rm \ion {He}{ii} \,\lambda 4686$/$\rm \ion {He}{i} \, \lambda 4713$ & $0.12\pm0.07$ \\		
		$\rm \ion {Si}{iii} \, \lambda 4552$/$\rm \ion {He}{ii} \, \lambda 4542$ & $1.59\pm0.18$ \\
		$\rm \ion {Si}{iv} \, \lambda 4089$/$\rm \ion {He}{i}\, \lambda 4026$ & $1.20\pm0.19$ \\		
\hline
\end{tabular}
\end{table}

\subsection{Stellar Parameters}

The absorption lines from the star are broadened, suggesting that it is rapidly rotating. We use the moderate resolution SOAR spectra to measure the projected rotational velocity $v_{\rm rot} \, \textrm{sin} \, i$ in the manner described by \citet{2014ApJ...788L..27S} in the region of the 4922 \AA\ line. We convolved the spectrum of the star HD 114122 taken with the same setup with kernels spanning a range of $v_{\rm rot}\, \textrm{sin} \, i$ and assuming a standard limb darkening law. Cross-correlation of the convolved spectra with the original star yields a relation between the input  $v_{\rm rot} \, \textrm{sin} \, i$ and the FWHM of the cross-correlation peak. Seven of the SOAR spectra had high enough S/N for this measurement, and we found a final value of $v_{\rm rot} \, \textrm{sin} \, i = 174\pm34$ km s$^{-1}$. The uncertainty is the standard deviation of the measurements, which is primarily determined not by the quality of the spectra but by true changes in the shape of the line, perhaps due to a time-variable wind or outflow from the star. Unfortunately, the lack of an appropriate standard star makes it impossible to repeat this analysis for the MIKE spectra. There are indeed variations in the profile of the 4922 \AA\ line in the MIKE spectra, consistent with the findings from the SOAR spectra. From this estimate of $v_{\rm rot}\, \sin(i)$, the radius from the SED (see below) and the orbital inclination $i$ from \citet{2019MNRAS.489.4705J}, the rotational velocity of the primary star is $v_{\rm rot}=247 \pm 48 \, \rm km \,s^{-1}$, which implies a rotation period of $P_{\rm orb}=4.85\pm1.11$~d.

To estimate the surface temperature ($T_{\rm{eff}}$) and surface gravity ($\log(g)$), we compare the high-resolution MIKE spectra with the TLUSTY OSTAR02 and BSTAR06 model atmospheres at $Z/Z_\odot=0.5$ \citep{2003ApJS..146..417L,2007ApJS..169...83L}. The model spectra are convolved using $v \, \textrm{sin} \, i = 174$ km s$^{-1}$, assuming a standard limb darkening law. We use the BSTAR06 grid for $T_{\rm{eff}}\leq30000\, K$ and the OSTAR02 grid for $T_{\rm{eff}}>30000\, K$. The BSTAR06 grid extends to $T_{\rm{eff}}=30000 \, K$ with a grid spacing of $\rm \Delta T_{eff}=1000 \, K$ and $\Delta \log(g)=0.25$. In comparison, the OSTAR02 grid begins at $T_{\rm{eff}}=27500 \, K$ with a grid spacing of $\rm \Delta T_{eff}=2500 \, K$ and $\Delta \log(g)=0.25$. 

\begin{figure*}

	\includegraphics[width=0.99\textwidth]{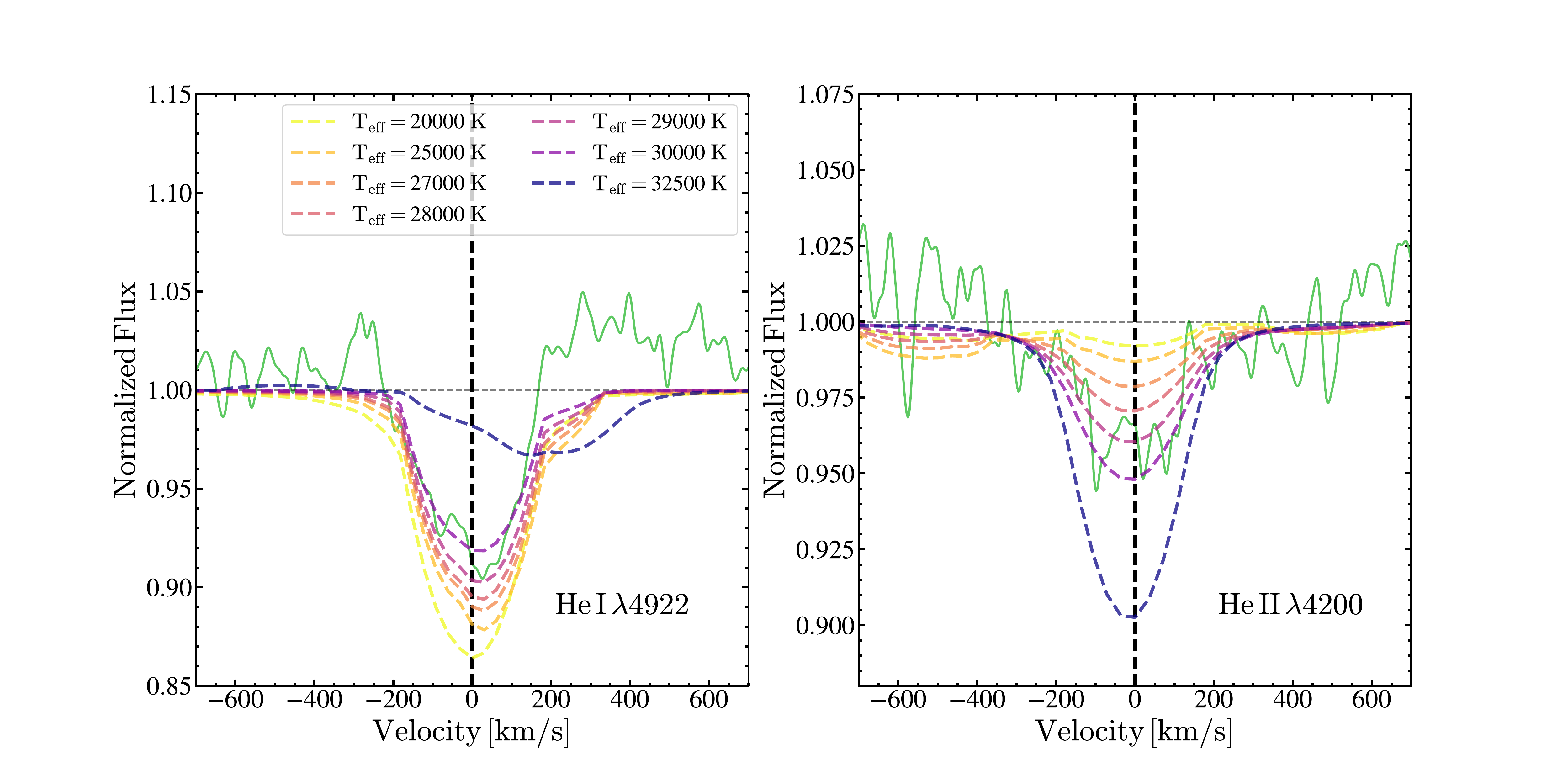}
    \caption{Comparison of various BSTARO6 and OSTAR02 model atmospheres \citep{2003ApJS..146..417L,2007ApJS..169...83L} at a fixed surface gravity of $\log(g)=3.25$ (dashed lines) against the MIKE spectrum at $\Phi=0.806$ for the $\ion {He}{i}\, \lambda 4922$ and $\ion {He}{ii}\, \lambda 4200$ absorption lines. The MIKE spectrum has been smoothed using a Gaussian kernel but not by enough to modify the line profiles.}
    \label{fig:fig4}

\end{figure*}

To determine $T_{\rm{eff}}$ and $\log(g)$, the MIKE spectra are compared with the model atmospheres at various $\ion{He}{i}$ and $\ion{He}{ii}$ absorption lines. The best overall fit is returned for a model atmosphere with $T_{\rm{eff}}=30000 \, K$ and $\log(g)=3.25$. Figure \ref{fig:fig4} compares the model atmospheres at a fixed surface gravity of $\log(g)=3.25$ but with various temperatures against the MIKE spectrum at $\Phi=0.806$ for the He lines $\ion {He}{i}\, \lambda 4922$ and $\ion {He}{ii}\, \lambda 4200$. We use these lines because they are less affected by emission from the disk. $\ion{He}{ii}$ is particularly sensitive to the surface temperature and the observed $\ion {He}{ii}\, \lambda 4200$ line best agrees with $T_{\rm{eff}}=30000$~K. Higher surface temperatures are clearly incompatible with our observations.

\begin{figure*}
	\includegraphics[width=0.90\textwidth]{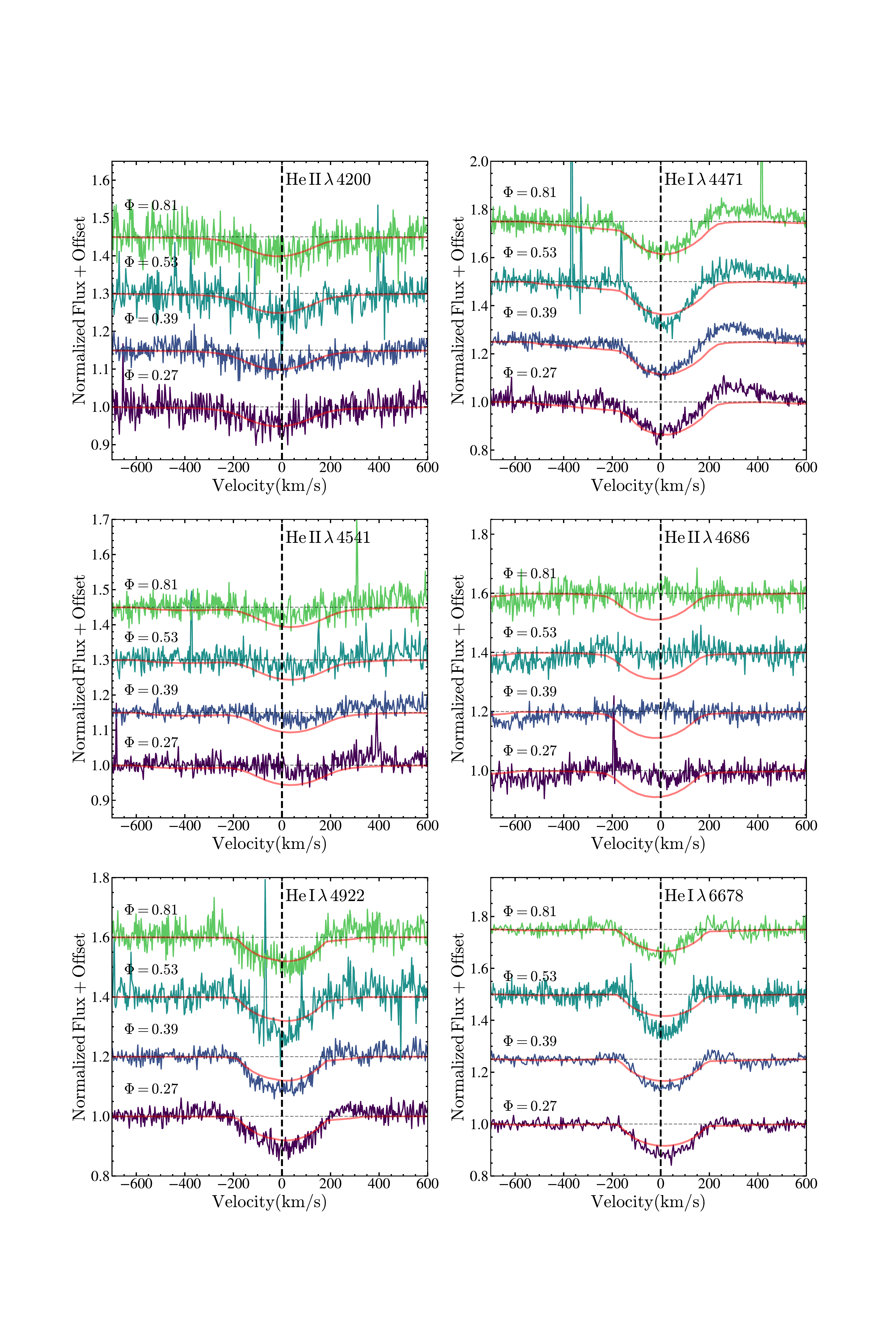}
	\vspace{-2cm}
    \caption{MIKE line profiles for $\ion {He}{ii}\, \lambda 4200$, $\ion {He}{i}\, \lambda 4471$, $\ion {He}{ii}\, \lambda 4541$, $\ion {He}{ii}\, \lambda 4686$, $\ion {He}{i}\, \lambda 4922$ and $\ion {He}{i}\, \lambda 6678$. The TLUSTY BSTAR06 model atmosphere with $T_{\rm{eff}}=30000 \, K$ and $\log(g)=3.25$ is shown in red \citep{2003ApJS..146..417L,2007ApJS..169...83L}.}
    \label{fig:fig5}

\end{figure*}

Figure \ref{fig:fig5} compares the best model atmosphere with $T_{\rm{eff}}=30000 \, K$ and $\log(g)=3.25$ to the MIKE spectra of selected $\ion{He}{i}$ and $\ion{He}{ii}$ absorption lines. There is clear evidence for spectral variability and emission components in the $\ion {He}{i}$ lines, particularly the emission component is distinctly seen in the $\ion {He}{i}\, \lambda 4471$ line. The $\ion {He}{ii}\, \lambda 4541$ and $\ion {He}{ii}\, \lambda 4686$ lines are relatively weak compared to the $\ion {He}{ii}\, \lambda 4200$ line. However, the equivalent width of the $\ion {He}{ii}\, \lambda 4686$ line grows weaker with increasing luminosity, and is weakest for the Ia luminosity class \citep{2018A&A...616A.135M}. This is entirely consistent with the B0 Iae spectral type that we assigned previously (see $\S3.1$).

Using the $UBV$ photometry in Table \ref{tab:phot}, we can derive a photometric temperature using the classical reddening free $Q$ parameter \citep{1958LowOB...4...37J,2012AJ....144..130B}, defined as $Q= (U-B)-X(B-V),$ with $X= E(U-B)/E(B-V).$ The $Q$ parameter can then be used to estimate the temperature using the calibration given by \citet{1989AJ.....97..107M},
\begin{equation}
     \log({T_{\rm{eff}}})=3.944-0.267Q+0.364Q^2.
	\label{eq:qteff}
\end{equation} Using a standard value ($X=0.72$) for the reddening ratio, we obtain $Q=-0.9791$, which implies $\log({T_{\rm{eff}}})=4.554$ and $T_{\rm{eff}}\simeq36000 \, K$. Using a similar method, \citet{2002ApJS..141...81M} estimated a higher temperature of $T_{\rm{eff}} \simeq 39,000$~K for this source. These photometric estimates of the temperature are considerably higher than our spectroscopic estimate, but they are less reliable. The lack of strong $\ion {He}{ii}$ absorption lines in the spectra also rule out this higher temperature estimate.  

We used DUSTY \citep{1997MNRAS.287..799I,2001MNRAS.327..403E} inside a Markov Chain Monte Carlo wrapper \citep{2015MNRAS.452.2195A} to fit the spectral energy distribution (SED) of MACHO 80.7443.1718 using the 18 photometric measurements spanning $3.6\mu$m thorough the UVM2 band (Table \ref{tab:phot}). We assumed foreground extinction due to $R_V=3.1$ dust \citep{1989ApJ...345..245C} and used \citet{2003IAUS..210P.A20C} model atmospheres for the star. Since MACHO 80.7443.1718 is located in the LMC, we assume a distance of $d_{\rm LMC}=50 \, \rm kpc$ \citep{2013Natur.495...76P}. Assuming minimum luminosity uncertainties of 10\% for each band, the fits have $\chi^2/N_{dof} \simeq 1$ at fixed $T_{\rm{eff}}$. While these models are adequate for determining the luminosity and extinction at fixed temperature, they are not reliable for determining a temperature (especially since they all lie on the Rayleigh-Jeans side of the SED). For $T_{\rm{eff}} \simeq 30,000$~K, we obtain $\log (L/L_\odot) = 5.61 \pm 0.04$ with $E(B-V) \simeq 0.39 \pm 0.02$ mag. 

\begin{figure}

	\includegraphics[width=0.5\textwidth]{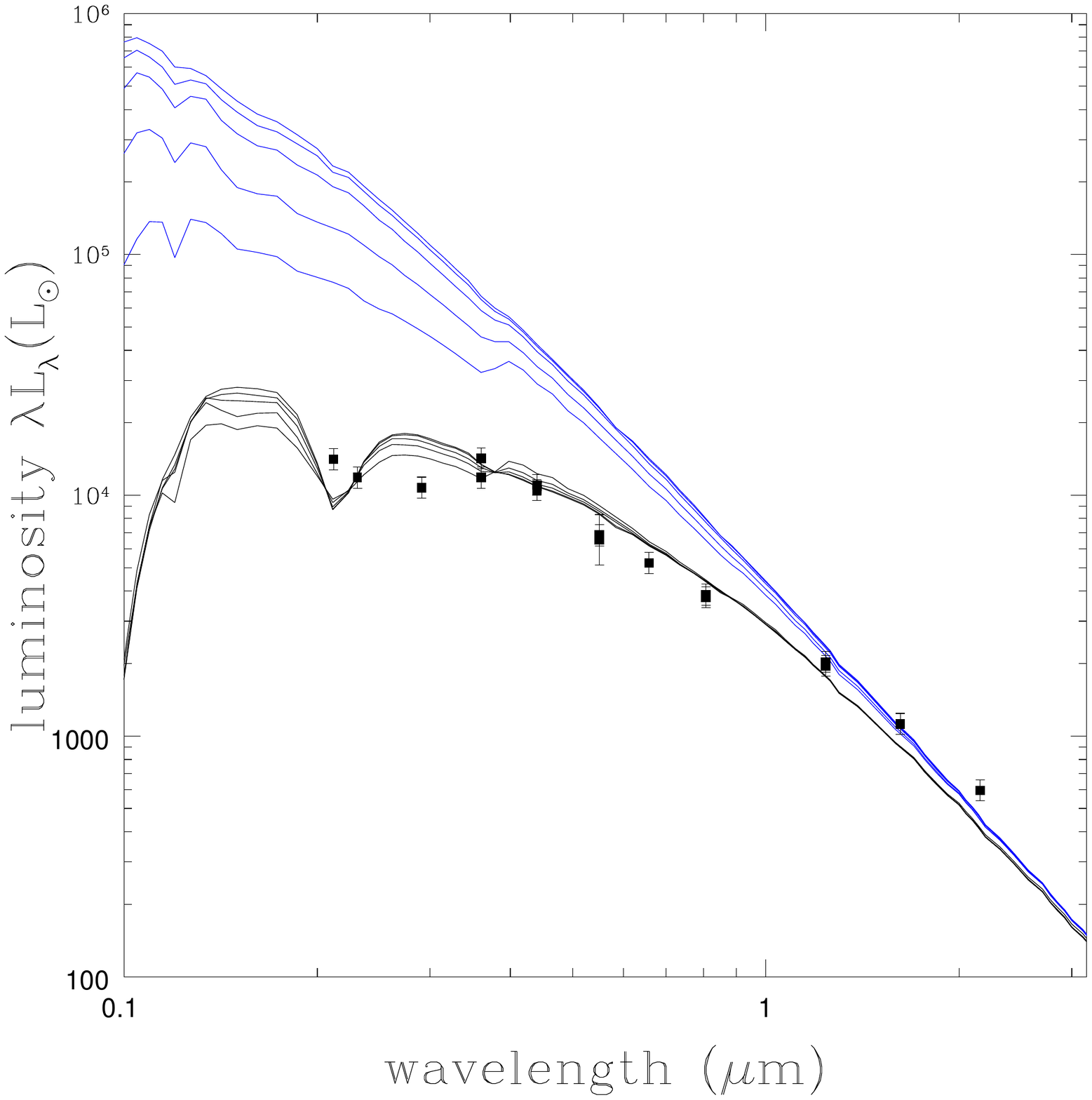}
	\vspace{-2cm}
    \caption{The spectral energy distribution (SED) for MACHO 80.7443.1718. The blue lines show the actual stellar SEDs at various $T_{\rm{eff}}$, and the black lines show these SEDs with reddening.}
    \label{fig:fig6}

\end{figure} 

\begin{figure}

	\includegraphics[width=0.5\textwidth]{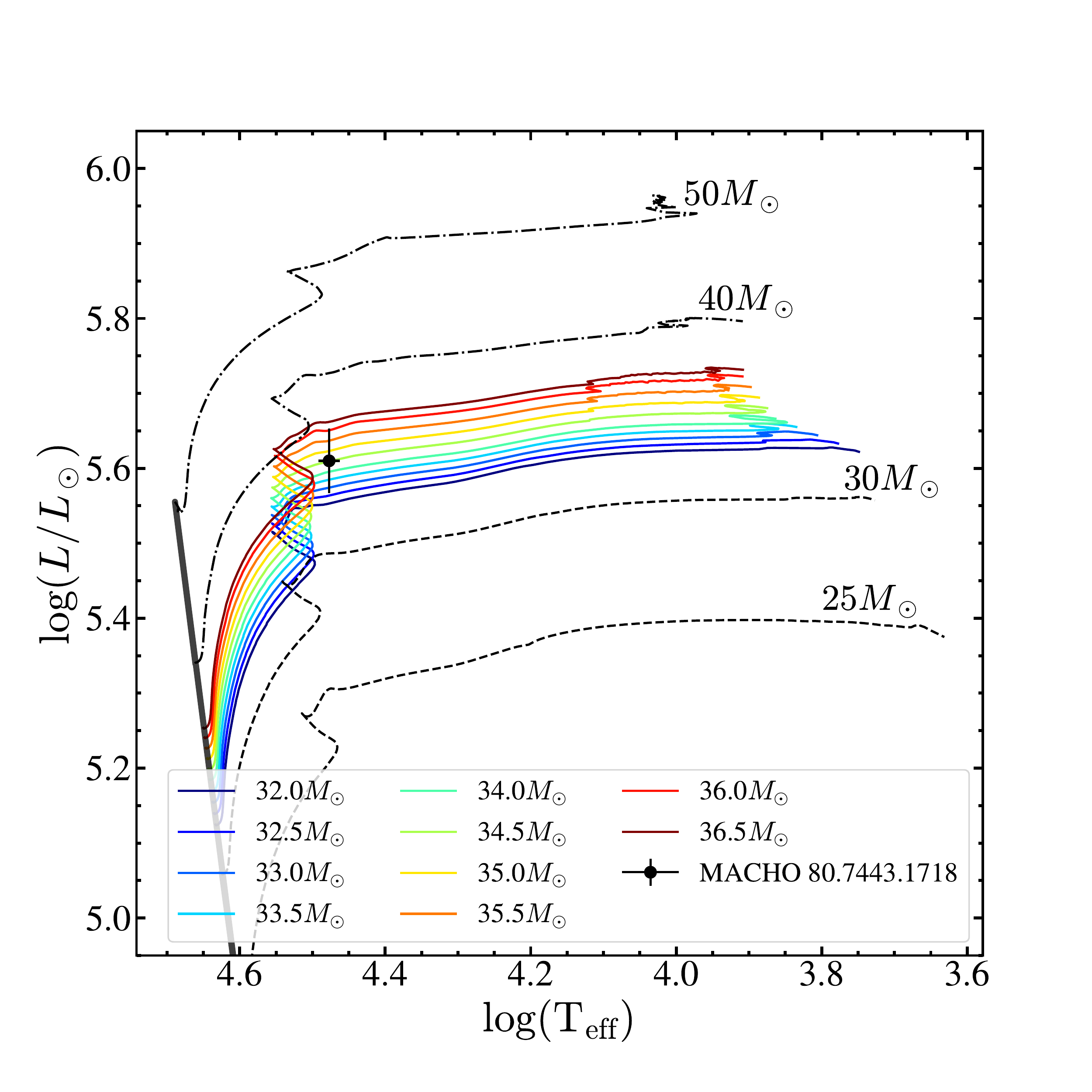}
    \caption{The Hertzsprung-Russell diagram (HRD) for MACHO 80.7443.1718. MIST evolutionary tracks \citep{2016ApJ...823..102C,2016ApJS..222....8D} are shown for stars at various masses with a metallicity of $\rm [Fe/H]=-0.4$.}
    \label{fig:fig7}

\end{figure} 

With the luminosity and temperature constrained, we can estimate the mass of the primary star using MIST stellar evolutionary tracks \citep{2016ApJ...823..102C,2016ApJS..222....8D} computed with the Modules for Experiments in Stellar Astrophysics code (MESA, \citealt{2011ApJS..192....3P,2013ApJS..208....4P,2015ApJS..220...15P}). We use tracks for stars with a metallicity of $\rm [Fe/H]=-0.4$, which is the average metallicity of the LMC \citep{2016MNRAS.455.1855C}. The position of MACHO 80.7443.1718 in the Hertzsprung-Russell Diagram (HRD) (Figure \ref{fig:fig7}) is consistent with a primary star of mass $M_1=34.5^{+1.5}_{-2.0} M_\odot$ starting to evolve across the Hertzsprung gap. In the MIST tracks, a $34.5 M_\odot$ star with $T_{\rm{eff}} \simeq 30,000$~K and $\log (L/L_\odot) \simeq 5.6$ has $\log(g)=3.2$ which is consistent with the spectroscopic estimate of $\log(g)=3.25 \pm 0.25$. Given the luminosity and the temperature, the radius of the primary is $R_1=23.7^{+2.9}_{-1.2} M_{\odot}$. For these estimates of the mass and radius, the breakup velocity of the star is ${\sim}530 \rm \, km\, s^{-1}$. This implies that MACHO 80.7443.1718 is rotating at $\Omega/\Omega_{\rm crit}{\sim}0.47$.

\begin{figure}

	\includegraphics[width=0.5\textwidth]{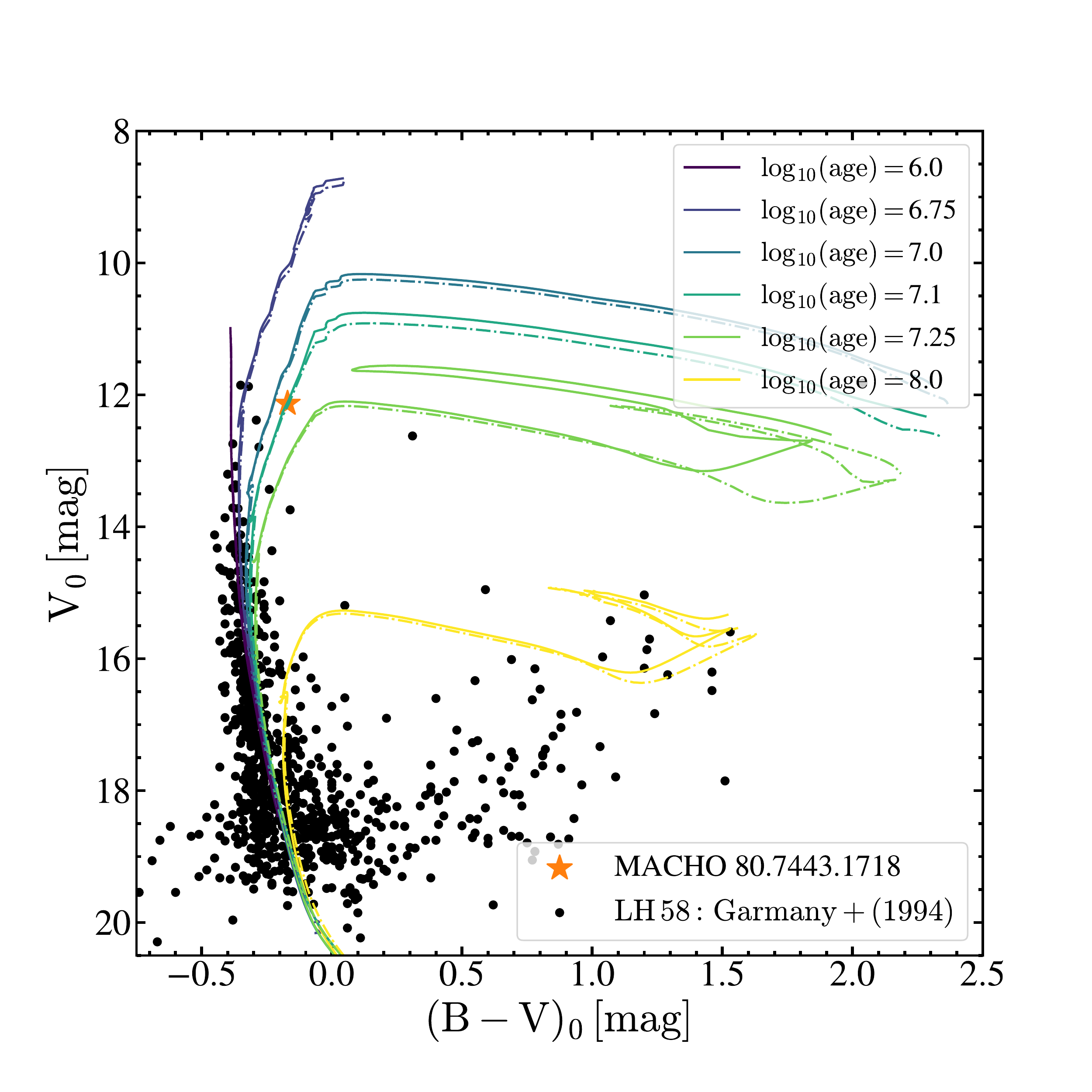}
    \caption{The de-reddened color-magnitude diagram for the LH58 OB association. MIST isochrones \citep{2016ApJ...823..102C,2016ApJS..222....8D} are shown for populations with a metallicity of $\rm [Fe/H]=-0.4$.}
    \label{fig:fig8}

\end{figure}

At its current evolutionary state in the MIST tracks, the age of the primary star is ${\sim}5.8$~Myr. Assuming that MACHO 80.7443.1718 is coeval with the LH58 OB association, we can verify its age by fitting isochrones to the LH58 color-magnitude diagram (CMD) using the photometry from \citet{1994AJ....108.1256G}. We estimate the global extinction towards LH58 by varying $E(B-V)$ to minimize the residuals between main sequence stars and a $1$~Myr MIST isochrone with $\rm [Fe/H]=-0.4$. We do not consider differential extinction to the members of this association. With this we obtain a reddening of $E(B-V)=0.17\pm0.01$ mag which we then use to de-redden the LH58 photometry. The reddening towards MACHO 80.7443.1718 ($E(B-V) \simeq 0.39 \pm 0.02$ mag) is larger than the reddening towards the LH58 association by $\Delta E(B-V) {\sim} 0.22$ mag, suggesting the presence of circumstellar dust consistent with the estimated age of the star. Figure \ref{fig:fig8} shows the de-reddened CMD for the LH58 OB association. The CMD is best-fit by an isochrone at ${\sim}5.6$~Myr, which is consistent with the age estimate of the primary from the MIST models (Figure \ref{fig:fig7}). We estimate the age of the LH58 association, and hence the age of MACHO 80.7443.1718 as $5.6^{+1.5}_{-1.1}$~Myr.

\subsection{Binary solution}

Previously, we used the analytical model from \citet{1995ApJ...449..294K} (their Equation 44) to model the flux variations produced by the tidal distortions at periastron \citep{2019MNRAS.489.4705J}. This model has been used to fit the light curves of other heartbeat stars \citep{2012ApJ...753...86T}. The fit returns the true anomaly, $\phi(t)$, the angle of periastron, $\omega$, the orbital inclination, $i$ and the eccentricity, $e$ as parameters. We attempted fitting a standard eclipsing binary model including irradiation and reflection effects using \verb"PHOEBE 2.3" \citep{2016ApJS..227...29P,2018ApJS..237...26H} but were unable to replicate the observed variability amplitude. To accurately model the photometric variability of this unique system, the circumstellar disk will also need to be considered in addition to modeling the dynamical tidal interactions and the two stellar components. We refit the combined \textit{TESS} light curve with the \citet{1995ApJ...449..294K} analytical model using the Monte Carlo Markov Chain sampler (MCMC) \verb"emcee" \citep{2013PASP..125..306F}. We obtain results that are consistent with our previous model (see \citealt{2019MNRAS.489.4705J}) with $i=43.9\degree \pm 0.2\degree$, $e=0.565 \pm 0.002$ and $\omega=298.2\degree \pm 1.0\degree$. Hereafter, we will use the orbital inclination from the updated fit. 

We fit eccentric Keplerian models to the radial velocities using the custom Monte Carlo sampler \verb"TheJoker" \citep{2017ApJ...837...20P}. Since the time span of the radial velocities is much shorter than the photometry, the orbital period is not left as a free parameter. Instead, we use a prior to constrain the period by the photometric estimate in Equation \ref{eq:asassnperiod} ($P = 32.8363\pm0.0085$ d). Leaving the rest of the parameters free, we find a BJD time of periastron $T_P = 2458505.595\pm0.301$ d, eccentricity $e = 0.506\pm0.033$, argument of periastron $\omega =  302.1\pm5.1^{\degree}$, semi-amplitude $K_1 =  62.9\pm4.0$ km s$^{-1}$, and systemic velocity $\gamma =  289.6\pm1.5$ km s$^{-1}$. There are no significant correlations between the variables except for the expected correlation between $T_P$ and $\omega$. This fit has a $\chi^2 = 10.0$ for 12 d.o.f.~and an rms residual of 4.6 km s$^{-1}$, and so is a good fit to the data.

\begin{figure*}

	\includegraphics[width=0.85\textwidth]{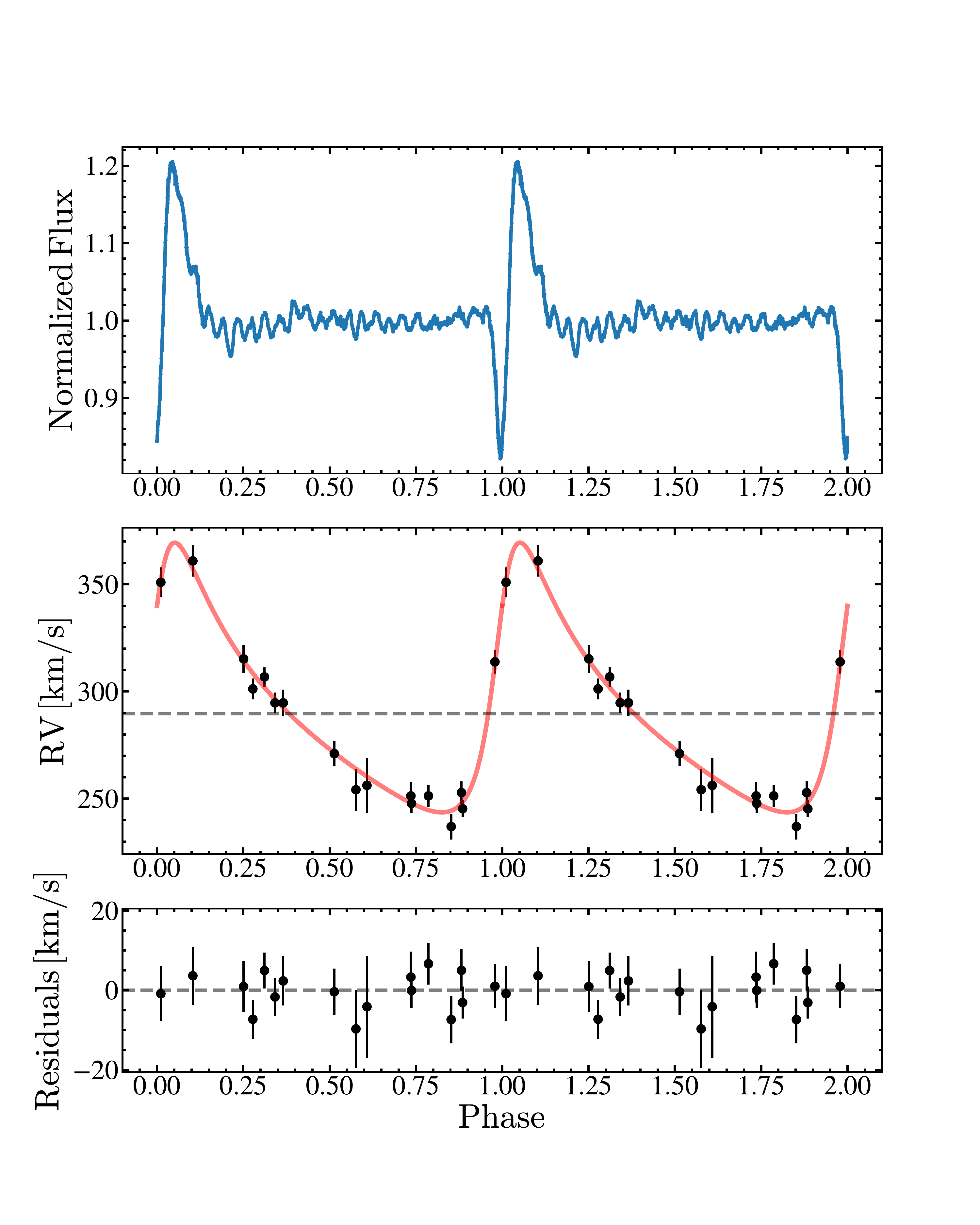}
    \caption{The best fitting Keplerian model to the radial velocity data (red line) for MACHO 80.7443.1718 (middle) compared to the averaged \textit{TESS} light curve (top). The bottom panel shows the velocity residuals.}
    \label{fig:fig9}

\end{figure*}

Fitting the TESS photometry gives an independent estimate of $T_P$ \citep{2019MNRAS.489.4705J}. This TESS value, extrapolated with the appropriate uncertainty to the first value within the time range of our velocities, is $T_P =  2458505.470\pm0.034$ d. This is entirely consistent with the radial velocity-only measurement of $T_P$, but more precise. If we re-fit the eccentric model with this estimate as a constraint, the values are very similar: $e = 0.507\pm0.033$, $\omega =  300.2\pm2.0^{\degree}$, $K_1 =  61.9\pm2.8$ km s$^{-1}$, and $\gamma =  289.4\pm1.5$ km s$^{-1}$. The goodness of fit is nearly identical at $\chi^2 = 10.3$ for 13 d.o.f.~and an rms of 4.7 km s$^{-1}$. This fit is shown in Figure \ref{fig:fig9}. We adopt these parameters as our final spectroscopic model. The value for $\omega$ derived from the radial velocities is entirely consistent with the photometric estimate to $<1\sigma$. The eccentricity derived from the radial velocities is slightly smaller (${\sim}1.8\sigma$) than that obtained from the fit to the light curve.

Using the posterior samples, the binary mass function is 
\begin{equation}
    f(M) = \frac{P_{\rm orb}K_1^3(1-e^2)^{3/2}}{(2 \pi G)} =  0.51^{+0.07}_{-0.06} M_{\odot}
\end{equation}

Using this value of $f(M)$, $M_1=34.5^{+1.5}_{-2.0} M_\odot$ and an inclination angle $i=43.9\pm0.2\degree$ from the heartbeat model, we find that the mass of the secondary is $M_2=15.7\pm 1.3 M_\odot$. 

\begin{figure*}

	\includegraphics[width=0.95\textwidth]{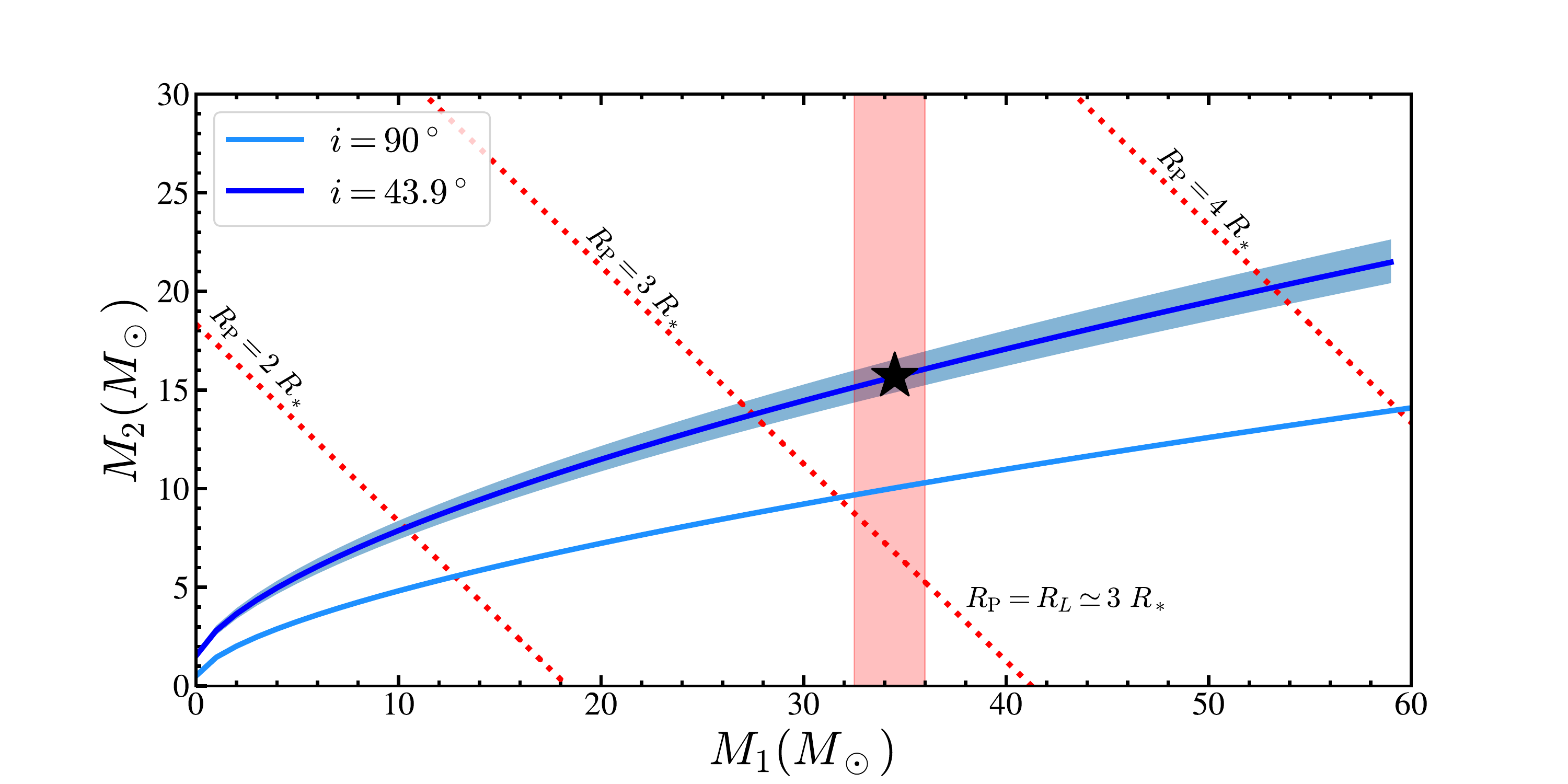}
    \caption{The mass of the companion ($M_2$) as a function of the mass of the primary ($M_1$). The binary mass function at inclinations of $i=90\degree$ and $43.9\degree$ are shown as the solid lines. Where the periastron distances would correspond to $R_P=2~R_*,~3~R_*,~4~R_*$ and the Roche Radius ($R_{L}\simeq3~R_*$) are shown by the dotted lines. The derived values for $M_1$ and $M_2$ are illustrated as the black star and the uncertainty in the mass of the primary star and the binary mass function are shaded.}
    \label{fig:figmp}

\end{figure*}

Figure \ref{fig:figmp} illustrates the mass of the companion as a function of the mass of the primary. Even for an edge-on orbit, the minimum mass of the companion is $M_2\simeq10 M_\odot$. The primary is well within its Roche radius ($R_L\simeq71~R_\odot\simeq3.0~R_*)$, but at periastron, it approaches Roche Lobe overflow with a fillout factor $f=R_*/R_{L,\rm ~peri}\simeq0.7$. We do not see strong evidence for a secondary star in our spectra. 

To obtain estimates of the stellar parameters of the secondary star, we use the MIST single star models assuming that the secondary is coeval with the primary. While these models do not account for binary interactions, they can provide a useful estimate of the properties of the unseen secondary star. The MIST models suggest that the secondary star has a similar temperature to the primary, with $T_{\rm{eff}}{\sim}31600\pm$~K. Unlike the primary, the secondary star should be on the main sequence with $\log(g)=4.1\pm0.1$. The MIST models indicate that the secondary has ${\sim}7\%$ of the primary's luminosity. Given that the secondary has a similar effective temperature and a considerably lower luminosity than the primary, it is challenging to recover the properties of the secondary star from the spectra. Using the calibrations from \citet{2005A&A...436.1049M} for main sequence stars, the secondary star should have an O9.5 spectral type. We will adopt this spectral type and the stellar properties from the MIST models for the secondary. The binary and stellar properties for MACHO 80.7443.1718 are summarized in Table \ref{tab:params}.

\begin{table*}
\begin{center}
\caption{Fundamental Parameters for MACHO 80.7443.1718}\label{tab:params}
	\begin{tabular}{r c c c c}
	\hline	
	& Parameter & Primary   &   Secondary & Reference  \\
	\hline
	\vspace{2mm}
	& $P_{\rm orb} (d)$ & \multicolumn{2}{c}{$32.83627\pm0.00846$  } & \citet{2019MNRAS.489.4705J} \\
	& $T_{P} (\rm{HJD}-2450000)$ & \multicolumn{2}{c}{$8505.470\pm0.034$} & \citet{2019MNRAS.489.4705J} \\
	& $i~(^\circ)$  & \multicolumn{2}{c}{$43.9\pm0.2 $ } & This work \\
	\vspace{2mm}
	& $\omega~(^\circ)$ & \multicolumn{2}{c}{$300.2\pm 2.0$} & This work\\
	\vspace{2mm}
	& $e$ & \multicolumn{2}{c}{$0.507\pm0.033$} & This work \\
	\vspace{2mm}
	& ${\gamma}~(\rm km s^{-1})$ &  \multicolumn{2}{c}{$289.4\pm1.5$} & This work \\
	\vspace{2mm}
	& $K~(\rm km s^{-1})$ & $61.9\pm2.8$ & -- & This work \\
	\vspace{2mm}	
	& $f(M)~(M_{\odot})$ &  \multicolumn{2}{c}{$0.51^{+0.07}_{-0.06}$} & This work \\
	\vspace{2mm}	
	& $v\, \sin(i)~(\rm km s^{-1})$ & $174\pm34$ & -- & This work \\
	\vspace{2mm}	
	& $T_{\rm{eff}}~(K)$ & $30000\pm1000$ & $31600\pm1000$& This work \\
	\vspace{2mm}
	& $\log (L/L_\odot)$ & $5.61\pm0.04$ & $4.46^{+0.11}_{-0.12}$ & This work \\
	\vspace{2mm}	
	& $\log(g)$ & $3.25\pm0.25$ & $4.1\pm0.1$& This work \\
	\vspace{2mm}	
	& $R~(R_{\odot})$ & $23.7_{-1.2}^{+2.9} R_{\odot}$ & $5.7_{-0.5}^{+0.4} R_{\odot}$& This work\\	
	\vspace{2mm}
	\vspace{2mm}
	& $M~(M_{\odot})$ & $34.5_{-2.0}^{+1.5} M_{\odot}$ & $15.7\pm 1.3 $ & This work\\
	\vspace{2mm}
	& $\rm [Fe/H]$ & \multicolumn{2}{c}{$-0.4$ (Fixed)} & This work \\
	\vspace{2mm}		
	& Spectral Type & B0 Iae & O9.5 V & This work \\
	\vspace{2mm}	
	& Age (Myr) & \multicolumn{2}{c}{$5.6_{-1.1}^{+1.5}$} & This work \\
	\vspace{2mm}	
	& $E(B-V)$~(mag) & \multicolumn{2}{c}{$0.39\pm0.02$} & This work \\
	\hline
	\vspace{2mm}	
\end{tabular}
\end{center}
\end{table*}

\begin{figure*}
	\includegraphics[width=0.9\textwidth]{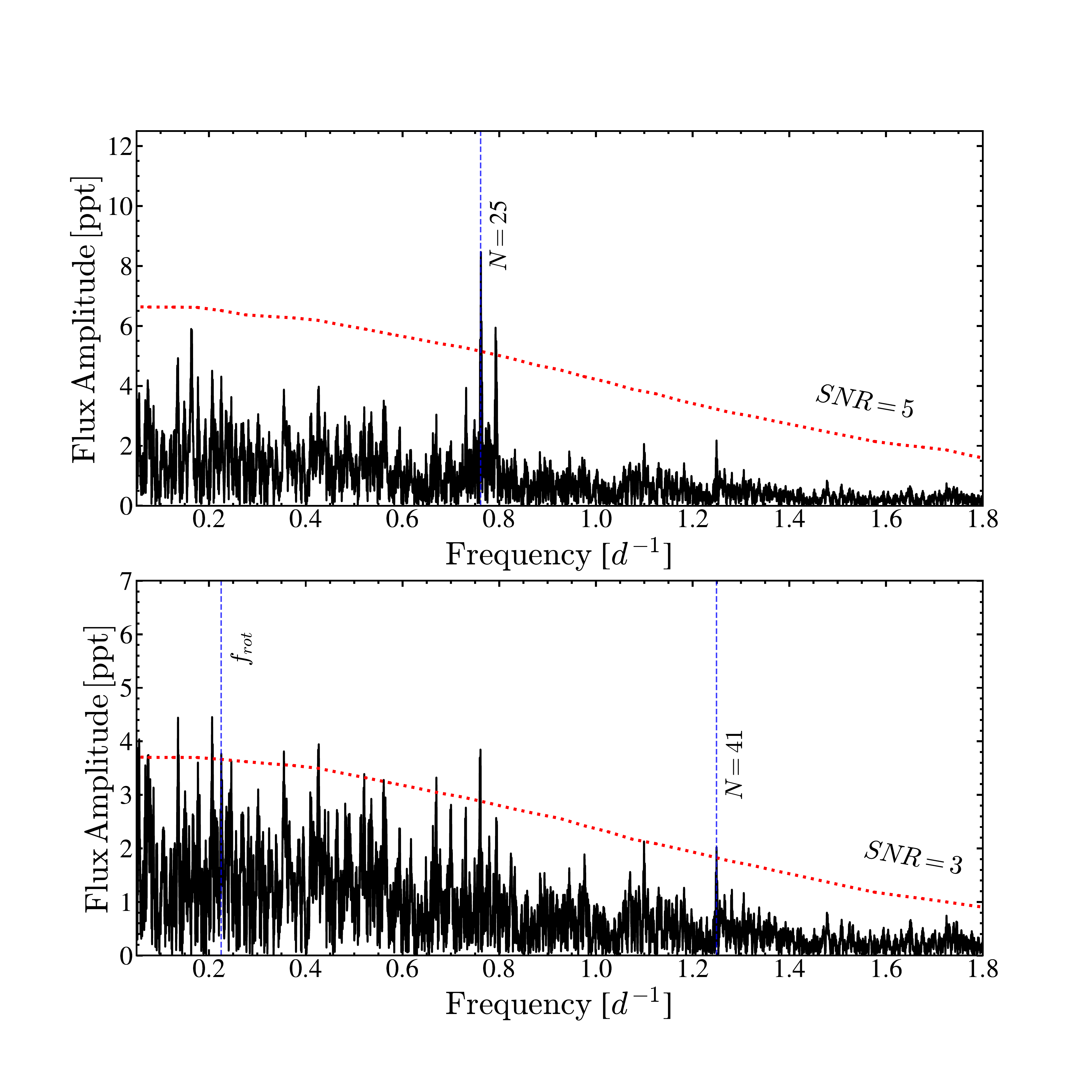}
    \caption{FFT power spectrum of the combined \textit{TESS} light curve before (top) and after (bottom) subtracting the dominant TEO at $N=25$. The red dashed lines show the flux amplitudes at which $SNR=5$ (top) and $SNR=3$ (bottom) as a function of frequency. The frequencies identified in our analysis are shown in blue.}
    \label{fig:fig10}
\end{figure*}

\subsection{Tidally Excited Oscillations}

\begin{table*}
	\centering
	\caption{Frequencies for MACHO 80.7443.1718, phased to periastron calculated using the multi-sector \textit{TESS} lightcurve. The errors in frequency, amplitude and phase are calculated through a Monte Carlo analysis.}
	\label{tab:orb}
    \begin{threeparttable}
\begin{tabular}{rrrrrrr}
		\hline
 		Type  & Frequency $\nu$ ($d^{-1}$) & Orbital Harmonic $N$ & $\nu/N\nu_{\rm orb}$ & Amplitude (ppt) & Phase & SNR\\
		\hline
		TEOs & & &\\
		& $0.76207 \pm 0.00001$ & 25 & 1.0009 & $8.4 \pm 0.2$ & $0.251 \pm 0.003$ & 8.2\\
		& $1.24944 \pm 0.00004$ & 41 & 1.0007 & $2.0 \pm 0.2$ & $0.969 \pm 0.014$ & 3.3\\
		\hline
		Rotation & & \\
		& $0.22530 \pm 0.00119$ & -- & -- & $4.4 \pm 0.8$ & $0.254 \pm 0.069$& 3.5\\			
\hline
\end{tabular}

  \end{threeparttable}      
\end{table*}

Previously, we analyzed both the ASAS-SN and TESS light curves for tidally excited oscillations (TEOs). In both the ASAS-SN and \textit{TESS} data, we identified a TEO corresponding to the $N=25$ orbital harmonic \citet{2019MNRAS.489.4705J}. We also recovered a possible TEO at $N=8$ and $N=7$ in the ASAS-SN and TESS data respectively. However, due to the limited baseline of the \textit{TESS} data (2 sectors) in our previous work, the resolution of the peaks in the Fast Fourier Transform (FFT) spectrum was poor. Here, we use all 18 sectors of \textit{TESS} data to analyze the FFT power spectrum for MACHO 80.7443.1718.

We search for TEOs in the \textit{TESS} data using only the orbital phases $0.25 \leq \Phi \leq 0.85$ by calculating the FFT using the \verb"Period04" software package \citep{2005CoAst.146...53L}. We iteratively whitened the light curves by fitting sinusoids using the dominant frequency. Peaks with signal-to-noise ratios ($\rm SNRs$) $>3$ were retained. Table \ref{tab:orb} lists the frequencies that were retrieved from this analysis of the \textit{TESS} data. The frequencies were optimized to reduce the light curve residuals and the uncertainties were determined using a Monte Carlo analysis within \verb"Period04". The TEO at the $N=25$ mode was the dominant frequency ($P=1.312$ d) with $\rm SNR=8.2$. Figure \ref{fig:fig10} illustrates the FFT power spectrum before and after subtracting the dominant TEO at $N=25$. Modes corresponding to the $N=22,\,23,\,24,\,26$ and $27$ orbital harmonics are also seen in the power spectrum prior to the subtraction of the TEO at $N=25$, but are not significant after the light curve is whitened. 

We identify two other modes in the FFT analysis at $P=4.438$~d and $P=0.800$~d. The latter mode corresponds to a TEO at the $N=41$ orbital harmonic, however the peak at $P=4.438$~d is significantly (${\sim}40\% $) different from the nearest orbital harmonic. A similar frequency was also recovered from the ASAS-SN and TESS data in \citet{2019MNRAS.489.4705J} with $P=4.148$~d and $P=4.621$~d, which we previously attributed to a TEO. However, this mode is very similar to the predicted orbital period of the primary star ($P_{\rm orb}=4.85\pm1.11$~d) that we estimated from the measurement of $v_{\rm rot} \, \textrm{sin} \, i$ in $\S3.2$, and is unlikely to be a TEO given its significant deviation from the nearest TEO. This period implies that $v_{\rm rot} \, \textrm{sin} \, i = 188\pm23$ km s$^{-1}$ assuming the radius from the SED fit ($\S 3.2$) and the orbital inclination from the heartbeat model ($\S 3.3$), which is consistent with the spectroscopic estimate ($v_{\rm rot} \, \textrm{sin} \, i = 174\pm34$ km s$^{-1}$). Studies of the photometric variability in Be stars have noted that periodic variability longward of 2 days could correspond to the beating of non-radial pulsation modes, stellar rotation, Rossby modes or circumstellar activity \citep{1996A&A...311..579S, 2017AJ....153..252L}. The position of MACHO 80.7443.1718 in the HRD (Figure \ref{fig:fig7}), and the period at $P=4.438$ d rules out pulsations \citep{2007MNRAS.375L..21M}. Given that this mode is within ${\sim}8\%$ of the expected rotational period of the primary, we believe that this signal originates due to the rotation of the star.

 We attempt to identify the pulsation modes corresponding to the $N=25$ and $N=41$ TEOs using tidal theory and models of stellar structure (see for, e.g., \citealt{2017MNRAS.472.1538F,2020ApJ...888...95G,Cheng2020}). Assuming that the TEOs are adiabatic and are standing waves that are not fine-tuned, \citet{2020ApJ...888...95G} approximates the pulsation phases for the dominant $l=2$ modes in heartbeat stars as
\begin{equation}
\label{phasesteos}
\phi_{l=2, m} =\left\{
\begin{array}{rcl}
0.25+m\phi_0  & & {\textrm{if}\  m=2 \ \ \textrm{or} -2}\\

0.25  & & {\textrm{if}\  m=0 }
\end{array} \right.
\end{equation}
where $\phi_0=0.25-\omega/2\pi$ is the observer's longitudinal coordinate and $m$ is the mode azimuthal number. These phases are defined with respect to the time of periastron ($\phi=0$). In the case of MACHO 80.7443.1718, $\phi_0\simeq-0.58$. The retrograde ($m=2$) or prograde ($m=-2$) modes have a 180$\degree$  phase ambiguity, thus a 0.5 phase offset can be added or subtracted from Equation \ref{phasesteos}. Therefore, the $m=|2|$ modes can have $\phi\simeq0.41$ or $\phi\simeq0.92$. For the $N=25$ TEO we found $\phi\simeq0.25$, which is entirely consistent with the $m=0$ mode ($\phi=0.25$). The $N=41$ TEO has $\phi\simeq1$, which is consistent with the $m=|2|$ mode. The phase deviation for the $N=41$ TEO $\Delta \phi (m=|2|) \simeq0.05$, can occur due to the non-adiabicity of the pulsations and spin-orbit misalignments \citep{2020ApJ...888...95G}. 

We also attempted to track the changes in amplitude and phase for the three frequencies in Table \ref{tab:orb} across the periastron passages in the \textit{TESS} data. Once it became clear that simple models of the oscillations would not work, we did not pursue this further. It would be interesting however, to look for variations in the individual modes across periastron passages.

\subsection{The B[e] phenomenon}

The presence of emission lines in both the SOAR and MIKE spectra suggest that MACHO 80.7443.1718 contains a circumstellar disk. This phenomenon is commonly associated with the classical Be stars \citep{2013A&ARv..21...69R}. Classical Be stars are rapidly rotating main-sequence B stars that form a Keplerian ``decretion'' disk \citep{2013A&ARv..21...69R}. However, this phenomenon also extends to supergiants like MACHO 80.7443.1718 \citep{2019Galax...7...83K}. B[e] supergiants have early B spectral types and show strong Balmer emission alongside narrow emission lines from both permitted and forbidden transitions \citep{2010A&A...517A..30K,2019Galax...7...83K}. B[e] supergiants (hereafter B[e] SGs) in the Magellanic Clouds typically have luminosities in the range $\log (L/L_\odot){\sim}4.0-6.0$ and have excesses in the near-infrared (NIR) caused by hot ${\sim}1000$~K dust \citep{1998A&A...340..117L,2006ASPC..355..135Z}. B[e] SG winds have a hot, fast line-driven polar component alongside a cool, slow, equatorial component from which forbidden emission lines (for e.g., [\ion{O}{i}] and [\ion{Fe}{ii}] lines) originate \citep{1998A&A...340..117L,2006ASPC..355..135Z, 2019Galax...7...83K}.

\begin{figure}
	\vspace{-1cm}	
	\includegraphics[width=0.48\textwidth]{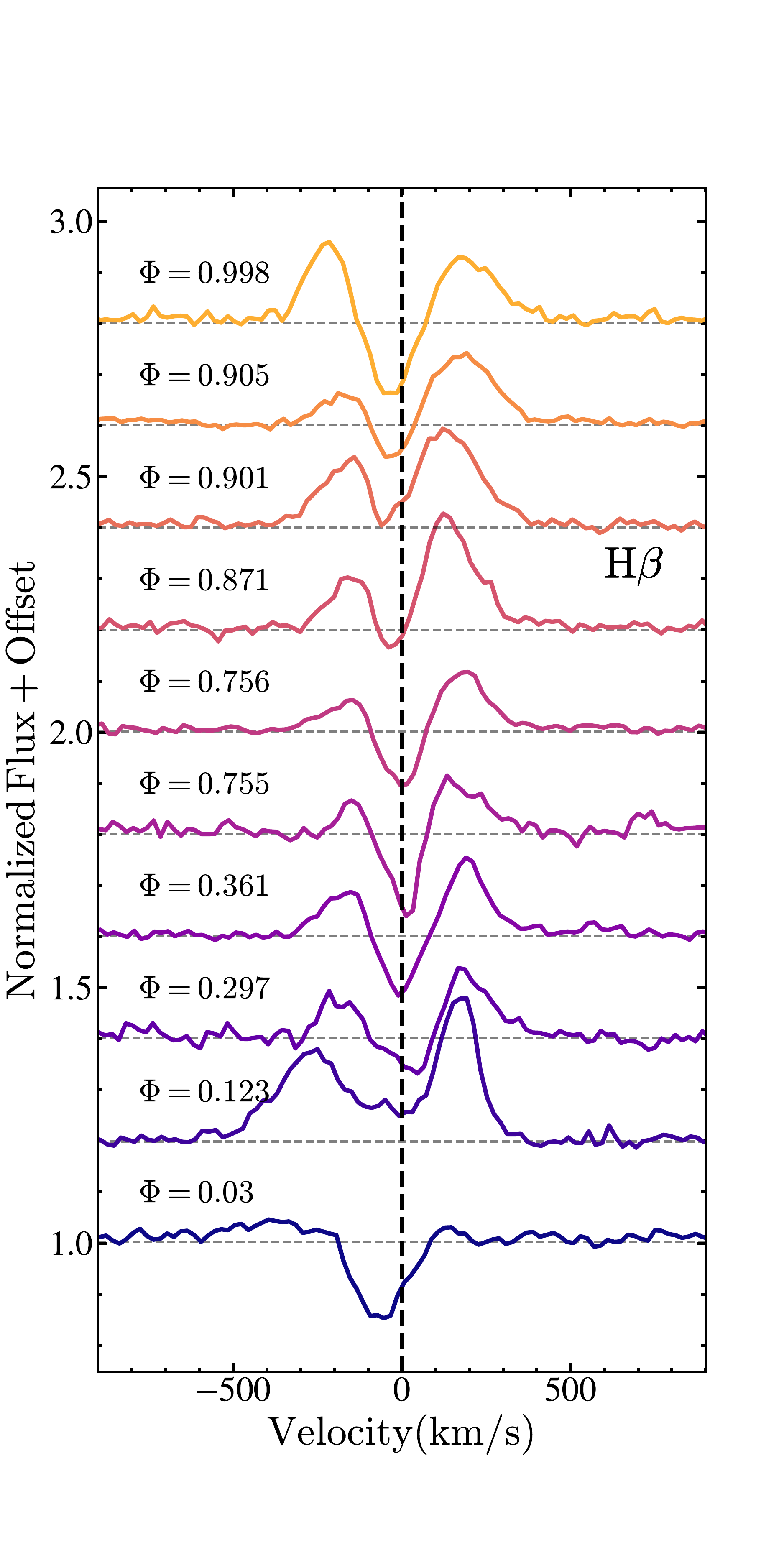}
	\vspace{-1.5cm}	
    \caption{SOAR $\rm H\beta$ line profiles for MACHO 80.7443.1718 sorted by orbital phase.}
    \label{fig:fig11}
\end{figure}

The Balmer line profiles of MACHO 80.7443.1718 show evidence of disk formation and disruption coinciding with the binary orbit. Figure \ref{fig:fig11} shows the $\rm H\beta$ line profiles in the SOAR spectra as a function of orbital phase. The typical Gaussian $FWHM$ of the $\rm H\beta$ line is ${\sim}675\, \rm km\,s^{-1}$. $\rm H\beta$ emission is typically produced close to the star, at a few $R_*$ \citep{2000A&A...356.1003N}. Close to periastron ($\Phi=0.030$), we see little evidence of a disk, with $\rm H\beta$ in absorption and very little emission. In the span of ${\sim}4$ days ($\Phi=0.123$), the $\rm H\beta$ line profile evolves into a double-peaked line profile. Be outbursts occur due to mass outflows from the stellar surface to the circumstellar disk \citep{2017AJ....153..252L}. At the intermediate orbital phases where TEOs dominate ($0.25 \leq \Phi \leq 0.85$), the $\rm H\beta$ line profile shows deep shell absorption with emission wings. At late phases, before periastron passage ($\Phi>0.85$), the shell-like line profile evolves to show emission, although the emission is weaker than that seen right after periastron. 

\begin{figure}
	\includegraphics[width=0.45\textwidth]{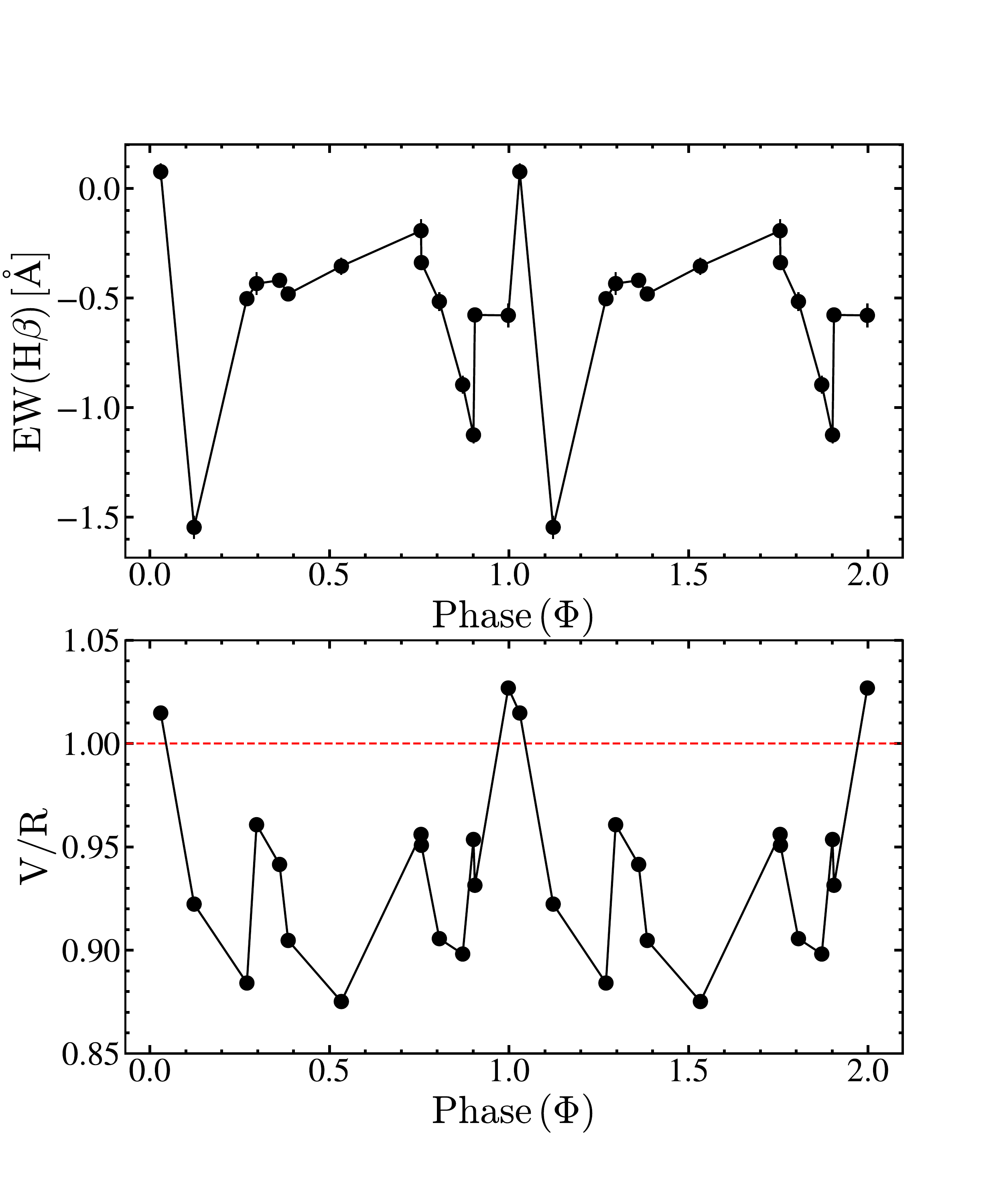}
	\vspace{-0.5cm}		
    \caption{The equivalent width of the $\rm H\beta$ line (top) and the V/R ratio (bottom) as a function of orbital phase $\Phi$.}
    \label{fig:figvrevol}
\end{figure}

The violet-to-red ($V/R$) ratio of the double-peaked emission lines is a useful quantity to study Be star emission lines \citep{1991A&A...241..159M}. The violet ($V$) peak dominates when material approaches the observer whereas the red ($R$) peak dominates when material moves away from the observer. Variations in the $V/R$ ratio are related to the dynamics of the cirumstellar disk which include global disk oscillations \citep{2017AJ....153..252L}. Figure \ref{fig:figvrevol} shows the evolution of the $V/R$ ratio and the equivalent widths ($\rm EW(H\beta)$) of the $\rm H\beta$ line profiles with the orbital phase. This system shows significant $V/R$ variability. At periastron, the $\rm H\beta$ symmetry is $V>R$, which immediately reverses to $V<R$ within $\Delta\Phi{\sim}0.2$. The equivalent width appears to be correlated with the orbital phase, and suggests that the system undergoes significant outbursts immediately before ($\Phi{\sim}0.9$) and after ($\Phi{\sim}0.1$) approaching periastron. At the intermediate orbital phases ($0.25 \leq \Phi \leq 0.85$), the equivalent widths are mostly similar.

\begin{figure*}
	\includegraphics[width=0.9\textwidth]{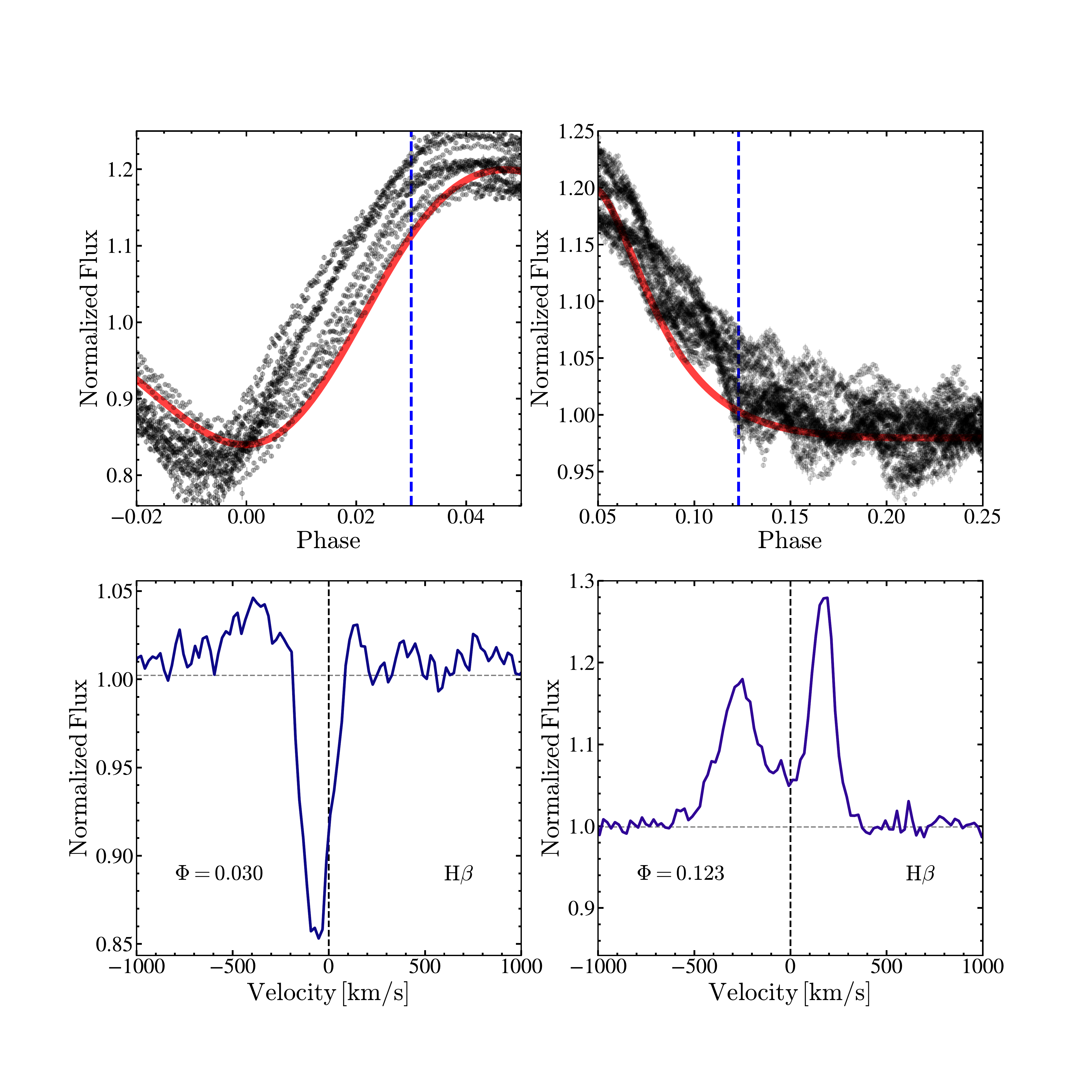}
	\vspace{-0.5cm}		
    \caption{The \textit{TESS} light curves (top) and SOAR $\rm H\beta$ line profiles (bottom) at $\Phi=0.030$ and $\Phi=0.123$ (vertical lines in top panels) corresponding to periastron passage and the first spectroscopic epoch with evidence of a circumstellar disk. The \citet{1995ApJ...449..294K} model for tidal distortions at periastron is shown in red.}
    \label{fig:fig12}
\end{figure*}

Figure \ref{fig:fig12} shows the \textit{TESS} photometry and the SOAR $\rm H\beta$ line profiles at $\Phi=0.030$ and $\Phi=0.123$ corresponding to periastron passage and the first spectroscopic epoch with evidence of a circumstellar disk respectively. When compared to a model of the tidal distortions at periastron \citep{1995ApJ...449..294K}, the \textit{TESS} light curve shows photometric evidence of outbursts occurring immediately following periastron while the star is tidally deformed. In fact, the scatter in the \textit{TESS} light curve beyond $\Phi{\sim}0.05$ suggests that the circumstellar disk might form earlier than $\Phi=0.123$. The TEOs following periastron passage likely contribute to mass loss and disk formation.

\begin{figure*}
	\includegraphics[width=1\textwidth]{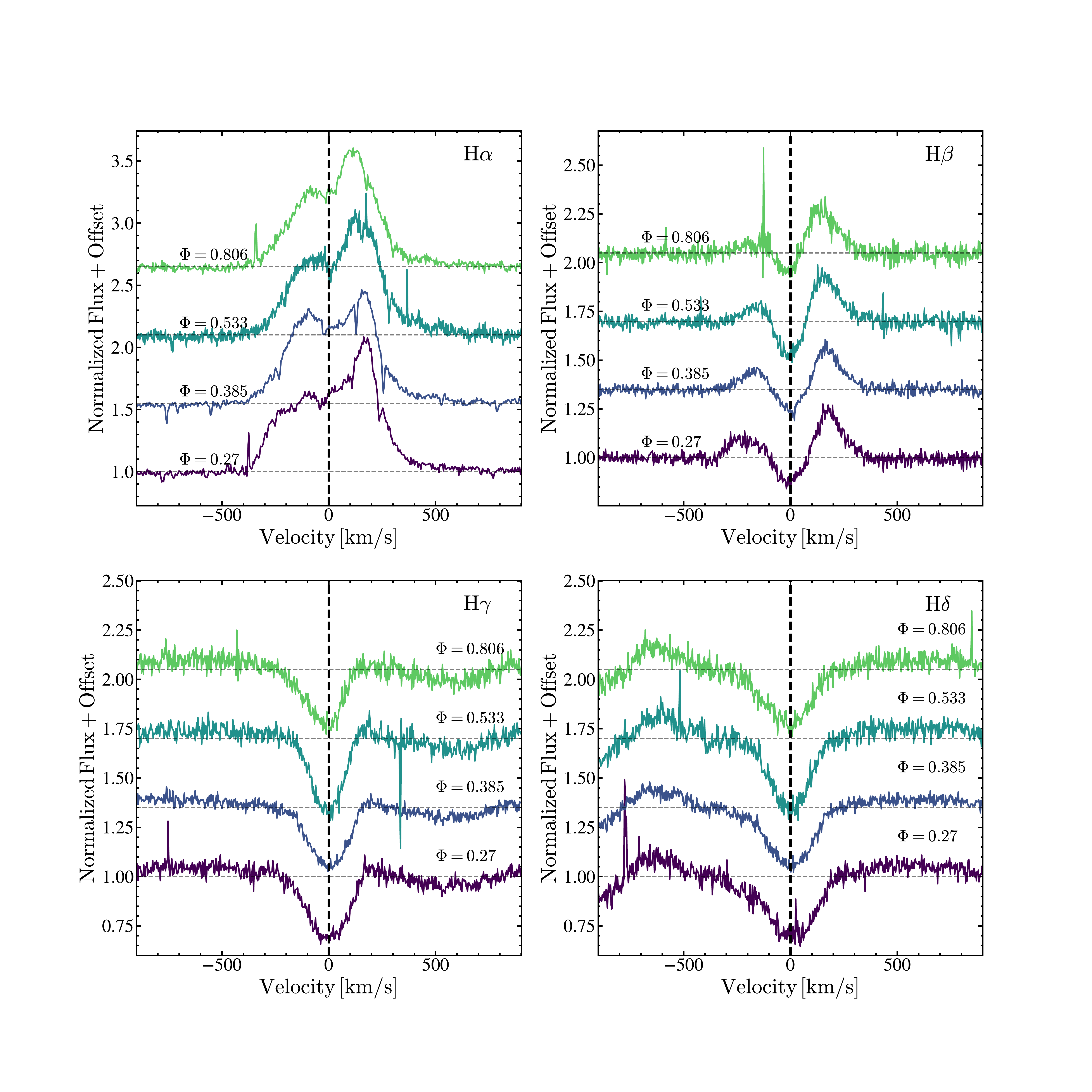}
	\vspace{-2cm}	
    \caption{MIKE line profiles for $\rm H\alpha$, $\rm H\beta$, $\rm H\gamma$ and $\rm H\delta$ sorted by phase.}
    \label{fig:fig13}
\end{figure*}

Figure \ref{fig:fig13} shows the Balmer line profiles in the high-resolution MIKE spectra as a function of orbital phase. There is clear evidence for strong $\rm H\alpha$ emission. Previous studies of Be stars have shown that $H\alpha$ emission probes a region up to ${\sim}10 R_*$ \citep{1995A&A...302..751H,1997ApJ...479..477Q}. The Gaussian $FWHM$ of the $\rm H\alpha$ line is ${\sim}620\, \rm km\,s^{-1}$ and its equivalent width $\rm EW(H\alpha)$ varies from ${\sim}14-26$~\angstrom. The $\rm H\alpha$ line profile is complex, with numerous sharp absorption/emission components that vary with orbital phase. The MIKE $\rm H\beta$ line profiles are comparable to those obtained from SOAR and show shell absorption alongside emission wings at intermediary orbital phases.

\begin{figure*}

	\includegraphics[width=0.9\textwidth]{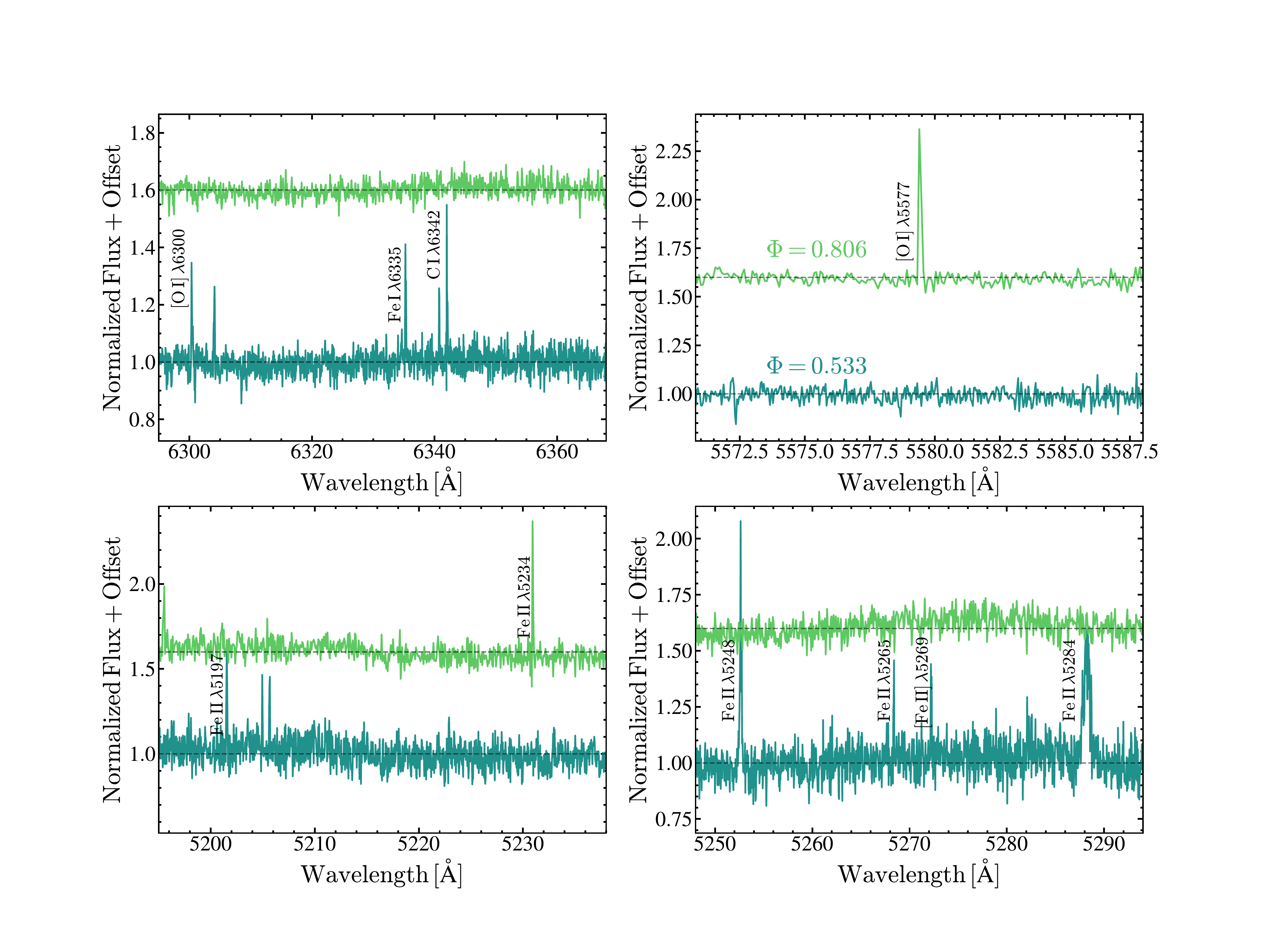}
	\vspace{-1cm}	
    \caption{MIKE spectra at $\Phi=0.533$ and $\Phi=0.806$ indicating the presence of \ion{O}{i}, \ion{C}{i}, \ion{Fe}{ii}, [\ion{O}{i}] and [\ion{Fe}{ii}] emission lines.}
    \label{fig:fig14}
\end{figure*}

To confirm the B[e] classification, we search for the presence of forbidden emission lines in the high-resolution MIKE spectra, focusing on the [\ion{O}{i}] and [\ion{Fe}{ii}] lines (Figure \ref{fig:fig14}). Forbidden emission lines are optically thin and collisionally excited. We do not see evidence for [\ion{O}{i}] emission lines in the two MIKE spectra at $\Phi=0.270$ and $\Phi=0.385$. We note the presence of the $[\ion {O}{i}]\, \lambda 6300$ line (single-peaked) at $\Phi=0.533$, but not at $\Phi=0.806$. Compared to the radial velocity of the primary, the $[\ion {O}{i}]\, \lambda 6300$ line has a radial velocity of ${\sim}+15\rm km\,s^{-1}$ and a Gaussian $FWHM$ of ${\sim}46\, \rm km\,s^{-1}$. In contrast, the $[\ion {O}{i}]\, \lambda 5577$ line (single-peaked) is only seen at $\Phi=0.806$. The $[\ion {O}{i}]\, \lambda 5577$ line has a radial velocity of ${\sim}+130\, \rm km\,s^{-1}$ and a Gaussian $FWHM$ of ${\sim}28\, \rm km\,s^{-1}$. Even though the $[\ion {O}{i}]\, \lambda 6300$ and $[\ion {O}{i}]\, \lambda 6364$ lines originate from similar regions in terms of temperature and density \citep{2010A&A...517A..30K}, we do not see the presence of the $[\ion {O}{i}]\, \lambda 6364$ line in any of our MIKE spectra. There are more emission lines at $\Phi=0.533$ than at $\Phi=0.806$. Here, we also see evidence for single-peaked permitted \ion{Fe}{ii} and \ion{C}{i} and forbidden [\ion{Fe}{ii}] lines in the MIKE spectra. For example, the forbidden $[\ion {Fe}{ii}]\, \lambda 5269$ line at ${\sim}~+185\rm km\,s^{-1}$ has a Gaussian $FWHM$ of ${\sim}24\, \rm km\,s^{-1}$. In contrast, the permitted $\ion {Fe}{ii}\, \lambda 5265$ line at ${\sim}+195\rm km\,s^{-1}$ has a Gaussian $FWHM$ of ${\sim}11\, \rm km\,s^{-1}$.

\subsection{Limit on X-rays}
Taken simultaneously with the \textit{Swift} UVOT observations presented in Section 2.1, MACHO 80.7443.1718 was also observed using the \textit{Swift} X-ray Telescope \citep[XRT][]{Burrows2005}. To place constraints on the presence of X-ray emission associated with this source, we reprocessed observations sw0045437001-002 and sw0045569001-002 using the \textit{Swift} \textsc{xrtpipeline} version 0.13.2 and standard filter and screening criteria. Using a 30 arcsec source region centered on  MACHO 80.7443.1718 and a 150 arcsec source-free background region located at $(\alpha_{J2000}, \delta_{J2000})$=(05:25:30.6787,-68:48:28.700), we find no significant X-ray emission arising from the source. To place deep constraints on the presence of X-ray emission, we combined all four observations using \textsc{xselect} version 2.4k. In this merged dataset, we derive a 3$\sigma$ upperlimit on the 0.3-2.0 keV count rate of $2.0\times10^{-3}$. Assuming $E(B-V)=0.17$ from Section 3.2, and using the relation from \citep{Liszt2014}, the Galactic column density towards MACHO 80.7443.1718 is $N_{H}\sim1.41\times10^{21}$ cm$^{-2}$, which is consistent with the Galactic column density along the line of sight ($N_{H,LOS} = 1.44\times10^{21}$ cm$^{-2}$, \citealt{2016A&A...594A.116H}). Assuming an absorbed powerlaw with a photon index of 2, we derive absorbed (unabsorbed) flux of $5.0\times10^{-14}$ erg cm$^{-2}$ s$^{-1}$ ($8.4\times10^{-14}$ erg cm$^{-2}$ s$^{-1}$), which corresponds to an absorbed (unabsorbed) X-ray luminosity of $1.5\times10^{34}$ erg s$^{-1}$ ($2.5\times10^{34}$ erg s$^{-1}$).

\clearpage

\section{Discussion}

MACHO 80.7443.1718 is by far the most extreme heartbeat discovered. In this work, we have shown that it is also one of the most massive heartbeat star systems thus known with a total mass of $M_{\rm tot}\sim50~M_\odot$. We note that \citet{Kolaczek-Szymanski2021} discovered that the likely hierarchical quadruple system HD 5980 in the Small Magellanic Cloud with $M_{\rm tot}\gtrsim150~M_\odot$ contained a heartbeat star, making this the most massive heartbeat star system known. In comparison, $\iota$ Ori, which consists of a binary system with a O9 III primary and a B1 III-IV secondary has a total mass of $M_{\rm tot}\sim37~M_\odot$ \citep{2017MNRAS.467.2494P}. $\epsilon$ Lupi is another massive heartbeat star system composed of two early-type main sequence B stars with $M_{\rm tot}\sim20~M_\odot$ \citep{2019MNRAS.488...64P}. MACHO 80.7443.1718 is also the only confirmed heartbeat star system consisting of a blue supergiant with a circumstellar disk.

We can understand the short disk lifetime with analytical arguments. The orbital semi-major axis from the period and Kepler's third law in terms of the stellar radius $R_* \simeq 24 R_\odot$ is
\begin{equation}
     a = 71 R_\odot \left( { M_1 + M_2 \over 10 M_\odot } \right)^{1/2} \simeq 159 R_\odot \simeq 6.7 R_*
\end{equation} 
where $M_1 \simeq 34.5 M_\odot$  and $M_2 \simeq 15.7 M_\odot$. The pericentric radius $R_p = a(1-e)$ is,
\begin{equation}
    R_p  \simeq 35 R_\odot \left( { M_1 + M_2 \over 10 M_\odot } \right)^{1/2} \simeq 78 R_\odot \simeq 3.3 R_*.
\end{equation}For $q=M_1/M_2\simeq2.2$, $R_L/a=0.448$, and the Roche limit is $R_L \simeq 71 R_\odot \simeq 3.0 R_*,$ \citep{1983ApJ...268..368E}. At periastron, the primary comes close to filling its Roche lobe ($R_{L,\rm ~peri}\simeq1.5~ R_*$) with a fillout factor $f=R_*/R_{L,\rm ~peri}\simeq0.7$. Material in a disk around the primary star will extend beyond the stellar photosphere and so, we should not be surprised to see mass transfer from the disk to the secondary star. 

We can use the $\rm H\beta$ line to derive the radius of the $\rm H\beta$ emitting region using the estimate of 
\begin{equation}
    {R_\beta \over R_*} {\simeq} \left( v\sin(i) v_{\rm crit} \over 350\Delta \lambda \right)^j {\simeq} \left( 261 \over \Delta \lambda \right),
\end{equation} from \citet{1991A&A...241..159M}, where $v_{\rm crit}{\sim}530 \rm \, km\, s^{-1}$, $\Delta \lambda$ is the half-peak separation in the $\rm H\beta$ line profile and $j\approx1$ is a rotation exponent. While it is not the true radius of the disk, $R_\beta$ can track its evolution with orbital phase. $R_\beta$ evolves from ${\sim}1.2\,R_*$ at $\Phi=0.123$, to ${\sim}2\,R_*$ at $\Phi=0.871$, indicating the growth of the disk after periastron passage. In the spectra taken before periastron at $\Phi=0.905$ and at $\Phi=0.998$, this drops to $R_\beta \sim 1.4\,R_*$, which suggests that the disk starts to shrink as the two stars approach each other. $\rm H\beta$ is in emission at $\Phi=0.998$ but at $\Phi=0.030$, $\rm H\beta$ is in absorption and shows little evidence for a disk (see Figure \ref{fig:fig11}). Thus, in the span of ${\sim}4$~hours the disk around MACHO 80.7443.1718 appears to dissipate. At periastron, a disk beyond a radius $R_d>R_{L,\rm ~peri}\simeq 1.5 R_*$ will result in mass transfer to the secondary. Given that the disks around Be stars are larger than a few $R_*$, it is very likely that mass transfer occurs at periastron in this system. However, this framework is complicated by the tidal interactions exerted on the disk by the secondary.

To obtain a rough estimate of the stability of the disk during a
pericentric passage, we used the approach of \citet{Kochanek1993}.  We set up
the stars at apocenter with their nominal masses,  put a ring of test
particles in circular orbits around the Be star, and then followed their
evolution through one orbit.  As a simple metric for the stability of
the disk as a function
of radius, we simply examined the maximum and minimum radii of the test
particles at the
end and the fraction still bound to the Be star. Material closer than
$30R_\odot$ likely can
remain in a disk.  For example, at $30R_\odot$ ring ends up with a
radial spread of $19.8$ to $36.4 R_\odot$ with all particles bound. At
$25 R_\odot$ the radial range shrinks to $22.6$ to $26.6 R_\odot$.  At
$35 R_\odot$, the particles are still bound but the radial range has
expanded to $4.6$ to $47.4 R_\odot$.  More distant rings are
significantly disrupted.  At $40$ and $45 R_\odot$, the outer radius is
$ \sim 280 R_\odot$ and the bound fractions are 83\% and 72\%,
respectively.  While crude, this experiment suggests that only the very
inner portions of the disk can survive a pericentric encounter.

Be stars in eccentric binaries can actually have increased mass loss rates due to binary interactions. At periastron, the secondary component can trigger outbursts in the Be star due to tidal forces that depend on the stellar properties of the Be star and the configuration of the binary orbit \citep{2017AJ....153..252L}. The role of binary interactions in triggering Be outbursts have been studied with the Be star $\delta$ Sco which is in a highly eccentric ($e=0.94\pm0.01$) orbit with $P_{\rm orb}\sim10.6$~years \citep{2003A&A...408..305M}. Spectroscopic observations of $\delta$ Sco at periastron passage suggest an increase in the mass loss rate at periastron \citep{2001A&A...377..485M}. It has been suggested that non-radial pulsations are amplified at periastron, resulting in an increased mass loss rate \citep{2001A&A...377..485M}. In the case of MACHO 80.7443.1718, we have significant evidence for high-amplitude tidally excited oscillations following periastron passage, along with photometric and spectroscopic evidence of outbursts immediately (${\sim}4$~days) following periastron (See Figure \ref{fig:fig12}). This provides strong evidence for amplified non-radial pulsations at periastron triggering mass loss in MACHO 80.7443.1718.

Given that both the components of this system are fairly massive and on a close orbit, this system maybe useful for studying the role of binary interactions in the formation of compact objects. We evolved a model of this system with the \verb"COSMIC" binary population synthesis model, assuming the default prescriptions in \citet{2020ApJ...898...71B}. The BPS models show that the primary will undergo Roche Lobe overflow as it evolves, increasing the mass of the secondary to ${\sim}24\, M_\odot$. Following mass transfer, the primary will end up on the He main-sequence with ${\sim}12\, M_\odot$. Subsequently, the primary undergoes a supernova explosion, resulting in the formation of a bound black hole of mass $M_{\rm BH}\sim8\, M_\odot$. The secondary evolves and forms a He main-sequence star with ${\sim}7\, M_\odot$. Following the supernova of the secondary star, it will form a neutron star of mass $M_{\rm NS}\sim1.6\, M_\odot$, and the binary is disrupted.

\section{Conclusions}

Using \textit{TESS} photometry and spectroscopic observations from MIKE and SOAR, we have studied the extreme LMC heartbeat star MACHO 80.7443.1718. We find that:

\begin{enumerate}
  \item MACHO 80.7443.1718 is a heartbeat star with the most extreme brightness variations observed to date. The ASAS-SN and \textit{TESS} light curves show photometric variations of ${\sim}40\%$ at periastron due to tidal distortions and variations of ${\sim}10\%$ due to tidally excited oscillations outside periastron.
  \item MACHO 80.7443.1718 is a massive binary on a short period ($P_{\rm orb}=32.83627\pm0.00846\,{\rm d}$),  eccentric ($e=0.507\pm0.033$) orbit, composed of a B0 Iae supergiant with $M_1 \simeq 34.5 M_\odot$ and a O9.5V secondary with $M_2 \simeq 15.4 M_\odot$. Thus, this system is among the most massive heartbeat systems yet discovered.
  \item Assuming that MACHO 80.7443.1718 is co-eval with the LH58 OB association or estimates from its properties at CMD models, it is ${\sim}6$~Myr old.
  \item MACHO 80.7443.1718 is likely a B[e] supergiant, showing emission in the Balmer lines and other permitted/forbidden emission lines from metals like O and Fe. However, we do not identify the strong NIR excess that is typical for these objects.
  \item MACHO 80.7443.1718 contains a circumstellar disk which disappears and then reforms at each periastron passage. We see significant $V/R$ variability in the Balmer emission line profiles during an orbit.
  \item The disk rapidly dissipates at periastron which could indicate mass transfer to the secondary, but re-emerges immediately after. 
  \item MACHO 80.7443.1718 shows evidence for tidally excited oscillations at $N=25$ and $N=41$ which likely correspond to the $(l,m)=(2,0)$ and $(l,m)=(2,|2|)$ modes respectively. We also retrieved a frequency that is very similar to the expected rotation period of the primary star.
  
\end{enumerate}

To fully understand the complex behavior of this system, detailed models that consider the interaction of the circumstellar disk with the stellar components are necessary. Our results have provided useful clues at understanding this unique binary and have shown that this system is not just unusual because of its unprecedented variability amplitudes but that it is peculiar in more ways than one. Therefore, MACHO 80.7443.1718 is an excellent astrophysical laboratory to study not just the binary interactions amongst massive stars, but also their evolution and mass-loss properties. 

\section*{Acknowledgements}

We thank Dr. Jim Fuller for useful discussions on this system and for comments on the manuscript.

ASAS-SN is supported by the Gordon and Betty Moore
Foundation through grant GBMF5490 to the Ohio State
University, and NSF grants AST-1515927 and AST-1908570. Development of
ASAS-SN has been supported by NSF grant AST-0908816,
the Mt. Cuba Astronomical Foundation, the Center for Cosmology 
and AstroParticle Physics at the Ohio State University, 
the Chinese Academy of Sciences South America Center
for Astronomy (CAS- SACA), the Villum Foundation, and
George Skestos. 

T. J. acknowledges support from the Ohio State Presidential Fellowship. KZS and CSK are supported by NSF grants AST-1515927, AST-1814440, and 
AST-1908570. JS acknowledges support from the Packard Foundation. J.T.H. was supported by NASA award 80NSSC21K0136. BJS is supported by NSF grants AST-1908952, AST-1920392, 
and AST-1911074.  Support for JLP is provided in part by the
Ministry of Economy, Development, and Tourism's Millennium Science 
Initiative through grant IC120009, awarded to The Millennium Institute 
of Astrophysics, MAS. EA acknowledges NSF award AST-1751874, NASA awards 11-Fermi 80NSSC18K1746, 13-Fermi 80NSSC20K1535, and 16-Swift 80NSSC21K0173, and a Cottrell fellowship of the Research Corporation. KCD acknowledges funding from the McGill Space Institute, the Natural Sciences and Engineering Research Council of Canada (NSERC), and the McGill Bob Wares Science Innovation Prospectors Fund.
Parts of this research were supported by the Australian Research Council Centre of Excellence for All Sky Astrophysics in 3 Dimensions (ASTRO 3D), through project number CE170100013.

Based on observations obtained at the Southern Astrophysical Research (SOAR) telescope, which is a joint project of the Minist\'{e}rio da Ci\^{e}ncia, Tecnologia e Inova\c{c}\~{o}es (MCTI/LNA) do Brasil, the US National Science Foundation’s NOIRLab, the University of North Carolina at Chapel Hill (UNC), and Michigan State University (MSU).

This research has made use of the VizieR catalogue access tool, CDS, Strasbourg, France. 
This research also made use of Astropy, a community-developed core Python package for 
Astronomy (Astropy Collaboration, 2013).

\section*{Data Availability}

The data underlying this article will be shared on reasonable request to the corresponding author.






\vspace{1cm}

\bibliographystyle{mnras}
\bibliography{refhb} 




\bsp	
\label{lastpage}
\end{document}